\documentclass[paper]{JHEP3}
\usepackage{epsfig}
\usepackage{amsmath}
\usepackage{amssymb}
\usepackage{amsbsy}
\usepackage{bbm}

\newcommand{\tmmathbf}[1]{\ensuremath{{\bf #1}}}
\newcommand{\tmop}[1]{\ensuremath{{\rm #1}}}

\def\beq{\begin{equation}}
\def\beqn{\begin{eqnarray}}
\def\eeq{\end{equation}}
\def\eeqn{\end{eqnarray}}

\newcommand\HERWIG{{\tt HERWIG}}

\newcommand\PYTHIA{{\tt PYTHIA}}

\newcommand\SISCONE{{\tt SISCONE}}
\newcommand\ALPGEN{{\tt ALPGEN}}

\newcommand\LambdaQCD{\Lambda_{\scriptscriptstyle QCD}}

\def\lq{\left[} 
\def\rq{\right]} 
\def\rg{\right\}} 
\def\lg{\left\{} 
\def\({\left(} 
\def\){\right)}

\newcommand\sss{\mathchoice%
{\displaystyle}%
{\scriptstyle}%
{\scriptscriptstyle}%
{\scriptscriptstyle}%
}

\newcommand\nplus{\oplus}
\newcommand\nminus{\ominus}

\newcommand\splus{{\sss \nplus}}
\newcommand\sminus{{\sss \nminus}}

\newcommand\splusminus{{\mathchoice%
{\vplusminus\displaystyle}%
{\vplusminus\scriptstyle}%
{\vplusminus\scriptscriptstyle}%
{\vplusminus\scriptscriptstyle}%
}}

\newdimen\hbigcirc
\newdimen\wbigcirc

\newdimen\figwidth

\newcommand\captskip{\vskip -0.7cm}

\newcommand\vplusminus[1]{%
\settoheight{\hbigcirc}{$#1\bigcirc$}%
\settowidth{\wbigcirc}{$#1\bigcirc$}%
\makebox[\wbigcirc]{%
\makebox[0pt]{\rule[0.4\hbigcirc]{0.5\wbigcirc}{0.05\hbigcirc}}%
\makebox[0pt]{\rule[0.1\hbigcirc]{0.5\wbigcirc}{0.05\hbigcirc}}%
\makebox[0pt]{\rule[0.1\hbigcirc]{0.05\wbigcirc}{0.6\hbigcirc}}%
\makebox[0pt]{$#1\bigcirc$}}%
}

\newcommand\kplus{k_\splus}
\newcommand\kminus{k_\sminus}
\newcommand\Kplus{K_\splus}
\newcommand\Kminus{K_\sminus}

\newcommand\as{\alpha_{\sss\rm S}}

\newcommand\pt{p_{\sss\rm T}}

\newcommand\kt{k_{\sss\rm T}}

\newcommand\stepf{\theta}

\newcommand\MCatNLO{{\tt MC@NLO}}

\newcommand\CF{C_{\sss\rm F}}
\newcommand\TF{T_{\sss\rm F}}

\newcommand\POWHEG{{\tt POWHEG}}

\newcommand\Rad{\Phi_{\rm rad}}

\newcommand\muF{\mu_{\sss\rm F}}

\newcommand\qb{\bar{q}}

\newcommand\xpm{x}

\preprint{
NSF-KITP-08-74\\
Bicocca-FT-08-9\\
}
\title{NLO vector-boson production \\
 matched with shower in {\tt\bf POWHEG}}
\author{Simone Alioli\\
  Universit\`a di Milano-Bicocca and INFN, Sezione di Milano-Bicocca\\
  Piazza della Scienza 3, 20126 Milan, Italy\\
  E-mail: \email{Simone.Alioli@mib.infn.it}}
\author{Paolo Nason\\
  INFN, Sezione di Milano-Bicocca,
  Piazza della Scienza 3, 20126 Milan, Italy\\
  E-mail: \email{Paolo.Nason@mib.infn.it}}
\author{Carlo Oleari\\
  Universit\`a di Milano-Bicocca and INFN, Sezione di Milano-Bicocca\\
  Piazza della Scienza 3, 20126 Milan, Italy\\
  E-mail: \email{Carlo.Oleari@mib.infn.it}}
\author{Emanuele Re\\
  Universit\`a di Milano-Bicocca and INFN, Sezione di Milano-Bicocca\\
  Piazza della Scienza 3, 20126 Milan, Italy\\
  E-mail: \email{Emanuele.Re@mib.infn.it}}
\abstract{
We present a next-to-leading-order
calculation of $W/Z$ production interfaced to shower Monte Carlo,
implemented according to the \POWHEG{} method. Finite width effects,
$Z/\gamma$ interference and angular correlations of decay products
are included. A detailed comparison with \MCatNLO{} and \PYTHIA{}
is carried out.
}

\begin{document}

\section{Introduction}
The {\POWHEG} method, first suggested in ref.~\cite{Nason:2004rx}, has been
successfully applied to $Z$ pair production~\cite{Nason:2006hfa},
heavy-flavour production~\cite{Frixione:2007nw} and $e^+ e^-$ annihilation
into hadrons~\cite{LatundeDada:2006gx}. In ref.~\cite{Frixione:2007vw} a
general description of the method was given, and in particular its
implementation within the Catani-Seymour (CS) subtraction
scheme~\cite{Catani:1997vz} and within the Frixione-Kunszt-Signer
(FKS)~\cite{Frixione:1995ms,Frixione:1997np} approach.

In this paper we present an implementation of the $W$ and $Z$ hadroproduction
cross section in the {\POWHEG} framework, using the CS subtraction formalism.
All next-to-leading-order (NLO) calculations used in {\POWHEG} until now have
been performed in the FKS method. In view of the popularity of the CS scheme,
we find desirable to explore more in detail its use within {\POWHEG}. In
ref.~{\cite{Frixione:2007vw}} an outline of the implementation of the
Drell-Yan production cross section in {\POWHEG} in the CS scheme was
given. In the present work we depart slightly from that approach. In
particular, we use a more appropriate form of the hardness variable used for
the generation of radiation. As a further point, for the case of $W$
production, if angular correlations in decay products are correctly taken into
account, a new problem arises. In fact, the Born-level $W$ cross section
vanishes when the fermion decay products are exactly in the opposite
direction of the incoming quark-antiquark pair, which causes a problem in the
generation of radiation within the \POWHEG{} method.  We show that this
problem has a simple solution, that can be easily generalized to all cases in
which the Born cross section vanishes.

The paper is organized as follows. In sec.~\ref{sec:description} we describe
how we performed the calculation for the NLO $W$ and $Z$ cross section. In
sec.~\ref{sec:powheg} we discuss the {\POWHEG} implementation and how to deal
with vanishing Born cross sections. In
sec.~\ref{sec:results} we show our results for several kinematic variables
and compare them with {\MCatNLO}~\cite{Frixione:2002ik} and
\PYTHIA{}~6.4~\cite{Sjostrand:2006za}.  Finally, in sec.~\ref{sec:conc}, we
give our conclusions.

\section{Description of the calculation}\label{sec:description}

\subsection{Kinematics\label{sec:kinematics}}
\subsubsection{Born kinematics}
We begin by considering the Born process for the annihilation of a quark and
an antiquark into a lepton-antilepton pair{\footnote{In case of $W$
    production the quark-antiquark and lepton-antilepton pairs have different
    flavour. We focus here for simplicity on leptonic decays of the vector
    bosons. Hadronic decays are treated similarly.}} $q + \bar{q} \rightarrow
l + \overline{l}$. Following ref. {\cite{Frixione:2007vw}}, we denote by
$k_{\splus}$ and $k_{\sminus}$ the incoming quark momenta, and by $k_1$ and
$k_2$ the outgoing fermion momenta. We call $K_\splus$ and $K_\sminus$ the
incoming hadron momenta and define the momentum fractions $x_\splusminus$ as
\begin{equation}
k_\splusminus=x_\splusminus K_\splusminus \;.
\end{equation}
We choose our reference frame with the $z$ axis along the
$k_{\splus}$ direction. We introduce the following variables
\begin{equation}
  M^2 = (k_1 + k_2)^2,\qquad\quad
 Y = \frac{1}{2} \log \frac{(k_1 + k_2)^0 + (k_1
  + k_2)^3}{(k_1 + k_2)^0 - (k_1 + k_2)^3} ,
\end{equation}
that characterize the invariant mass and rapidity of the virtual vector
boson.{\footnote{The virtuality of the lepton pair $M^2$ will be distributed
according to a Breit-Wigner formula around the squared mass of the vector
boson $M_{V^{}}^2$ (where $V$ stands for either the $W^{\pm}$ or the $Z$).}}
We also introduce the angle $\theta_l$ that represents the angle between the
outgoing lepton and the $k_{\splus}$ momentum, in the centre-of-mass frame of
the lepton pair. The azimuthal orientation of the decay products is
irrelevant here, since the cross sections do not depend upon
it. We thus fix it to zero. At the end of the generation of the event,
we perform a uniform, random azimuthal rotation of the whole event, in order
to cover all final-state phase space. The set of variables $M^2$, $Y$ and
$\theta_l$ fully parametrize our Born kinematics. From them we can
reconstruct
\begin{equation}
  x_{\splus} = \sqrt{\frac{M^2}{S}} e^Y, \qquad\quad
  x_{\sminus} = \sqrt{\frac{M^2}{S}}
  e^{- Y},
\end{equation}
where $S=(\Kplus + \Kminus)^2$.  The leptons' momenta are first reconstructed
in the longitudinal rest frame of the lepton pair, where each lepton has
energy equal to $M/2$ and where the lepton momentum forms an angle $\theta_l$
with the $\splus$ direction and has zero azimuth (i.e.~it lies in the $z, x$
plane and has positive $x$ component). The leptons' momenta are then boosted
with boost angle $Y$.

The Born phase space in terms of these variables is written as
\begin{equation}
  d \tmmathbf{\Phi}_2 = dx_\splus \, dx_\sminus (2\pi)^4 \delta^4(k_\splus+
k_\sminus -k_1 -k_2) \frac{d^3 k_1}{(2\pi)^3 2 k_1^0}\,
\frac{d^3 k_2}{(2\pi)^3 2 k_2^0}
=\frac{1}{S}  \frac{1}{16 \pi} d M^2 \, d Y \, d \cos
  \theta_l \, \frac{d\phi_l}{2\pi}\,.
\end{equation}
\subsubsection{Real-emission kinematics}
The real emission process is described by the final-state momenta $k_1$, $k_2$
and $k_3$, where $k_1$ and $k_2$ have the same meaning as before, and $k_3$ is
the momentum of the radiated light parton.
In the \POWHEG{} framework, applied in the context of the CS
subtraction method, one introduces a different real phase-space parametrization
for each CS dipole. In the present case, we have two CS dipoles, with the
two incoming partons playing the role of the emitter and the spectator.
We consider the case of the $\splus$ collinear direction.
Thus, the emitter is the incoming parton with momentum $k_\splus$.
We introduce the variable
\begin{equation}
  \xpm = 1 - \frac{(k_{\splus} + k_{\sminus}) \cdot
  k_3}{k_{\splus} \cdot k_{\sminus}},
\end{equation}
and the momenta
\beqn
  K &=& k_1 + k_2 = k_{\splus} + k_{\sminus} - k_3\\
  \bar{K} &=& \xpm \, k_{\splus} + k_{\sminus} \,.
\eeqn
Observe that $K^2 = \bar{K}^2$, which is the condition that fixes the value of
$\xpm $. When $k_3$ is collinear to $k_{\splus}$ we
have
\begin{equation}
  \xpm k_{\splus} = k_{\splus} - k_3,
\end{equation}
and $K = \bar{K}$. Following ref.~\cite{Catani:1997vz}, we introduce the
boost tensor
\begin{equation}
  \Lambda^{\mu}_{\phantom{\mu}\nu} (K, \bar{K}) =
  g^{\mu}_{\phantom{\mu}\nu} - \frac{2 (K + \bar{K})^{\mu}
  (K + \bar{K})_{\nu}}{(K + \bar{K})^2} + \frac{2 \bar{K}^{\mu}
  K_{\nu}}{K^2}\, ,
\end{equation}
the barred momenta
\begin{equation}
  \bar{k}^{\mu}_r = \Lambda^{\mu}_{\phantom{\mu}\nu} (K,
  \bar{K}) \, k^{\nu}_r \hspace{2em} \hspace{2em} r = 1, 2,
\end{equation}
the barred-momentum fractions
\begin{equation}\label{eq:barredout}
  \bar{x}_{\splus} = \xpm \, x_{\splus}, \qquad\quad
\bar{x}_{\sminus} =  x_{\sminus} \,,
\end{equation}
and the barred incoming momenta
\begin{equation} \label{eq:barredin}
  \bar{k}_{\splus} = \xpm k_{\splus} = \bar{x}_{\splus}
  K_{\splus}, \qquad\quad
\bar{k}_{\sminus} = k_{\sminus} = \bar{x}_{\sminus} K_{\sminus} \,.
\end{equation}
The barred momenta characterize the underlying-Born kinematics. We define then
\begin{equation}
  \bar{M}^2 = ( \bar{k}_1 + \bar{k}_2)^2 = (k_1 + k_2)^2, 
\qquad\quad
\bar{Y}_\splus =
  \frac{1}{2} \log  \frac{( \bar{k}_1 + \bar{k}_2)^0 + ( \bar{k}_1 +
  \bar{k}_2)^3}{( \bar{k}_1 + \bar{k}_2)^0 - ( \bar{k}_1 + \bar{k}_2)^3},
\end{equation}
and the angle $\bar{\theta}_l$ is defined as in the Born case, but in term of
the momenta $\bar{k}_{\splus}$, $\bar{k}_{\sminus}$, $\bar{k}_1$ and
$\bar{k}_2$.

The radiation variables are given by
\begin{equation}
  \xpm,\qquad\quad v = \frac{k_{\splus} \cdot k_3}{k_{\splus}
  \cdot k_{\sminus}},\qquad\quad \phi,
\end{equation}
where $\phi$ is the azimuth of $k_3$ around the $z$ direction.

From the set of variables $\bar{M}^2$, $\bar{Y}_\splus$, $x$, $v$ and $\phi$
we can reconstruct the full production kinematics for the real-emission cross
section. We summarize the reconstruction procedure from
ref.~{\cite{Frixione:2007vw}}.
From $\bar{M}^2$ and $\bar{Y}$ we reconstruct the barred momenta, as for the
Born kinematics case. Then we reconstruct immediately
\begin{equation}
  k_{\splus} = \frac{\bar{k}_{\splus}}{x},\quad\quad  k_{\sminus} = \bar{k}_{\sminus},
\end{equation}
and then
\begin{equation}\label{eq:ktdef}
  k_3 = v k_{\sminus} + (1 - x - v) k_{\splus} + k_T,
\end{equation}
where $k_T$ has only transverse components. Its magnitude is determined by the
on shell condition $k_3^2 = 0$, which yields
\begin{equation}
\label{eq:ktsq}
  k_T^2 = 2 k_{\splus} \cdot k_{\sminus} (1 - x - v) v
\end{equation}
and its azimuth is $\phi$. We then construct the vectors
\begin{equation}
  K = k_{\splus} + k_{\sminus} - k_3,\qquad \quad
\bar{K} = x k_{\splus} + k_{\sminus},
\end{equation}
and the inverse boost
\begin{equation}
  \Lambda_{\mu \nu}^{- 1} (K, \bar{K}) = g_{\mu \nu} - \frac{2 (K +
  \bar{K})_{\mu} (K + \bar{K})_{\nu}}{(K + \bar{K})^2} + \frac{2 K_{\mu}
  \bar{K}_{\nu}}{K^2},
\end{equation}
from which we can compute the leptons' momenta
\begin{equation}
  k_r = \Lambda^{- 1} (K, \bar{K}) \,\bar{k}_r,\quad\quad  r = 1, 2 .
\end{equation}
The real-emission phase space can be expressed in a factorized form in terms
of the underlying Born kinematics phase space and of the radiation variables
\begin{equation}
  d {\bf \Phi}_3 = d {\bar{\bf \Phi}}_2 \, d \Phi_{\tmop{rad}},
\end{equation}
with
\begin{equation}
\label{eq:dphirad}
  d \Phi_{\tmop{rad}} = \frac{\bar{M}^2}{16 \pi^2} \, \frac{d \phi}{2 \pi}
  \,d v \,\frac{d x}{x^2} \,\theta (v)\, \theta \!\left( 1 - \frac{v}{1 - x}
  \right) \theta (x (1 - x)) \, \theta (x - \bar{x}_{\splus})
\end{equation}
and
\begin{equation}
  d {\bar{\bf \Phi}}_2 = \frac{1}{S}\,  \frac{1}{16 \pi} \,d \bar{M}^2\, d
  \bar{Y}\, d \cos \bar{\theta}_l \,.
\end{equation}
The kinematic variables corresponding to the $\sminus$ collinear direction
are reconstructed in full analogy. Observe that the underlying-Born variables
and the radiation variables depend in general upon the collinear region that
we are considering.  In the present case, while $\bar{M}$, $x$ and $\phi$ are
obviously independent of the region we are considering, $\bar{Y}$,
$\bar{\theta}_l$ and $v$ do depend upon it.  In order to avoid a too heavy
notation, we have refrained from appending $\splus$ or $\sminus$ indices to
the underlying Born and radiation variables.  When necessary, we will
put a $[\phantom{a}]_\splusminus$ ``context'' bracket around a formula,
meaning that the underlying Born and radiation variables inside it should
refer to the $\splusminus$ direction.

\subsection{Cross sections}
We have used the helicity amplitude method of
refs.~\cite{Hagiwara:1985yu,Hagiwara:1988pp} in order to compute the cross
sections including the vector-boson decay products.  For the $W$-boson
propagator we have taken
\begin{equation}\label{eq:wprop}
  \frac{- g_{\mu \nu} + q_{\mu} q_{\nu} / M_W^2}{q^2 - M_W^2 + i \Gamma_W
    M_W}
\end{equation}
and for the $Z/\gamma$-boson propagators, multiplied by the corresponding
couplings, 
\begin{equation}\label{eq:Zgamprop}
  g_l\, g_q \,\frac{- g_{\mu \nu} + q_{\mu} q_{\nu} / M_Z^2}{q^2 - M_Z^2 + i
  \Gamma_Z M_Z} + e_l\, e_q \,\frac{- g_{\mu \nu}}{q^2},
\end{equation}
where $g_l$, $g_q$ are the lepton and
quark couplings to the $Z$ (for given helicities), and $e_l$, $e_q$ are their
electric charges.

Following ref.~{\cite{Frixione:2007vw}}, we introduce the Born
$\mathcal{B}_{q\bar{q}}$ and the real-emission cross sections
$\mathcal{R}_{q\bar{q},g}$, $\mathcal{R}_{g\bar{q},q}$ and $\mathcal{R}_{q
g,\bar{q}}$, that represent the contributions for quark-antiquark,
gluon-antiquark and quark-gluon initiating processes. Notice that the flavour
of the outgoing particle in the subscript of $\mathcal{R}$ is also taken to
be incoming.  In the case of $Z$ production, $q$ and $\bar{q}$ are conjugate
in flavour. For $W^\pm$ production, because of flavour mixing, $q$ and
$\bar{q}$ may refer to different flavour species.  We thus assume that, in
general, $q$ and $\bar{q}$ may both represent any flavour, but, in general,
if $q$ is a quark, $\bar{q}$ is an antiquark, and viceversa.  $\mathcal{B}$
and $\mathcal{R}$ are obtained by taking the absolute value squared of the
corresponding helicity amplitude, summing over the helicities and colours of
the outgoing particles, averaging over the helicities and colour of the
initial partons, and multiplying by the flux factor $1 / (2 s)$ (see
eq.~(\ref{eq:kin_variables})). The soft-virtual term in the CS approach is
given  by (see eq.~(2.107) in ref.~{\cite{Frixione:2007vw}})
\begin{equation}
  \mathcal{V}_{q \bar{q}} = \frac{\alpha_S}{\pi} C_F  \mathcal{B}_{q
  \bar{q}}\;. 
\end{equation}
Defining
\begin{equation}
\label{eq:kin_variables}
 s = (\kplus+\kminus)^2, \quad\quad u = (k_{\splus} - k_3)^2=-s\,v,\quad\quad
 t = (k_{\sminus} - k_3)^2=-(1-x-v)\,s,
\end{equation}
the CS subtraction terms are given by
\begin{equation}
{\cal C}^\splus_{q\qb,g} = \left[-\frac{1}{u}
2\,g_s^2\,\CF\lg \frac{2}{1-\xpm} -\(1+\xpm\)\rg
{\cal B}_{q\bar{q}}(\bar{M},\bar{Y},\bar{\theta}_l)\right]_\splus\,,
\end{equation}
for gluon radiation from a $q\qb$ initial-state, and
\begin{equation}\label{eq:Cgqb}
{\cal C}_{g\qb,q} = \left[-\frac{1}{u}
2\,g_s^2\,\TF \lg 1-2\,\xpm\(1-\xpm\)\rg
{\cal B}_{q\bar{q}}(\bar{M},\bar{Y},\bar{\theta}_l) \right]_\splus\,,
\end{equation}
for the $g \qb$. Analogous formulae apply for the $q\bar{q}$ and the $qg$
counterterms in the $\sminus$ collinear direction.

The collinear remnants are given by
\begin{eqnarray}
\mathcal{G}_{\splus}^{q\qb,g}({\bf \Phi}_{2,\splus}) 
&=& \frac{\as}{2\pi}\CF\!\!\lq 
\( \frac{2}{1-z}\log\frac{(1-z)^2}{z}\)_+ 
- (1+z)\log\frac{(1-z)^2}{z} +(1-z) \right.
\nonumber\\
&&\phantom{\frac{\as}{2\pi}\CF\Big[\!\!}
\left.+\(\frac{2}{3}\pi^2-5\) \delta(1-z)+
\(\frac{1+z^2}{1-z}\)_+\log\frac{M^2}{\muF^2}\rq 
\left[{\cal B}_{q\bar{q}}(\bar{M},\bar{Y},\bar{\theta}_l)\right]_\splus\!,
\\
\mathcal{G}_{\splus}^{g\qb,q}({\bf \Phi}_{2,\splus})
 &=& \frac{\as}{2\pi}\TF \lg\hspace{-1mm}
\lq z^2+(1-z)^2\rq \hspace{-1.5mm}\lq \log\frac{(1-z)^2}{z}
+\log\frac{M^2}{\muF^2}\rq \hspace{-1.mm} +2z(1-z) \rg \!\!
\left[{\cal B}_{q\bar{q}}(\bar{M},\bar{Y},\bar{\theta}_l)\right]_\splus \! .
\nonumber\\
\end{eqnarray}
The ${\bf \Phi}_{2,\splus}$ notation, according to
ref.~\cite{Frixione:2007vw}, represents the set of variables
\begin{equation}\label{eq:collpluskin}
{\bf \Phi}_{2,\splus}=\{x_\splus,x_\sminus,z,k_1,k_2\},\qquad\quad
z\, x_\splus K_\splus +x_\sminus K_\sminus=k_1+k_2\,.
\end{equation}
We also associate an underlying Born configuration $\bar{\bf \Phi}_2$ to the
${\bf \Phi}_{2,\splus}$ kinematics, defined by
\begin{equation}
\bar{k}_\splus=z \,x_\splus K_\splus,\qquad \bar{k}_\sminus=x_\sminus
K_\sminus, \qquad \bar{k}_1=k_1,\quad \bar{k}_2=k_2\,.
\end{equation}
The other two collinear remnants, $\mathcal{G}_{\sminus}^{q\qb,g}({\bf
  \Phi}_{2,\sminus})$ and $\mathcal{G}_{\sminus}^{qg,\bar{q}}({\bf
  \Phi}_{2,\sminus})$, are equal to $\mathcal{G}_{\splus}^{q\qb,g}({\bf
  \Phi}_{2,\splus})$ and $\mathcal{G}_{\splus}^{g\qb,q}({\bf
  \Phi}_{2,\splus})$ respectively, with $\left[{\cal
    B}_{q\bar{q}}(\bar{M},\bar{Y},\bar{\theta}_l)\right]_\splus$ replaced by
$\left[{\cal B}_{q\bar{q}}(\bar{M},\bar{Y},\bar{\theta}_l)\right]_\sminus$.
We then introduce the notation $B$, $V$, $R$, $C$, $G$, to stand for
${\mathcal{B}}$, ${\mathcal{V}}$, ${\mathcal{R}}$, ${\mathcal{C}}$,
${\mathcal{G}}$, each multiplied by its appropriate parton densities.  The
differential cross section, multiplied by some infrared safe observable $O$,
can then be written as
\begin{eqnarray}
\langle O\rangle &=&\sum_{q\bar{q}}\Bigg\{
\int d{\bf \Phi}_2
 \left[B_{q\bar{q}}({\bf \Phi}_2) +V_{q\bar{q}}({\bf \Phi}_2)\right]
O({\bf \Phi}_2) \nonumber \\
&+& \int d {\bf \Phi}_3 \left\{
R_{q\bar{q},g}({\bf \Phi}_3) O({\bf \Phi}_3)
-C^\splus_{q\bar{q},g}({\bf \Phi}_3) \left[ O(\bar{\bf \Phi}_2) \right]_\splus
-C^\sminus_{q\bar{q},g}({\bf \Phi}_3) \left[ O(\bar{\bf \Phi}_2)
  \right]_\sminus \right\}
\nonumber \\
&+& \int d {\bf \Phi}_3 \left\{
R_{g\bar{q},q}({\bf \Phi}_3) O({\bf \Phi}_3)
-C_{g\bar{q},q}({\bf \Phi}_3) \left[ O(\bar{\bf \Phi}_2) \right]_\splus
\right\}
\nonumber \\
&+& \int d {\bf \Phi}_3 \left\{
R_{qg,\bar{q}}({\bf \Phi}_3) O({\bf \Phi}_3)
-C_{qg,\bar{q}}({\bf \Phi}_3) \left[ O(\bar{\bf \Phi}_2) \right]_\sminus
\right\}
\nonumber \\
&+&\int d {\bf \Phi}_{2,\splus} \left[
G_\splus^{q\bar{q},g}({\bf \Phi}_{2,\splus})
+
G_\splus^{g\bar{q},q}({\bf \Phi}_{2,\splus})\right]\,O({\bf \Phi}_{2,\splus})
\nonumber \\
&+&\int d {\bf \Phi}_{2,\sminus} \left[
 G_\sminus^{q\bar{q},g}({\bf \Phi}_{2,\sminus})
+G_\sminus^{qg,\bar{q}}({\bf \Phi}_{2,\sminus})\right]\,O({\bf
  \Phi}_{2,\sminus}) 
\Bigg\}\;.
\end{eqnarray}

\section{{\POWHEG} implementation}
\label{sec:powheg}
The starting point of a \POWHEG{} implementation is the inclusive cross
section at fixed underlying-Born flavour and kinematics. For the soft-virtual
and Born contributions the underlying Born kinematics is obviously given by
the Born kinematics itself.  For the collinear remnant, for example, in the
$\splus$ direction (see eq.~\ref{eq:collpluskin}) the underlying Born
kinematics is given by
\begin{equation}
\bar{\bf \Phi}_2=\{zx_\splus,x_\sminus, k_1,k_2\}\,.
\end{equation}
For the CS counterterms, the underlying Born kinematics is given by the
corresponding $\bar{\bf \Phi}_2$ variables defined in
eqs.~(\ref{eq:barredout}) and~(\ref{eq:barredin}). In order to assign an
underlying Born kinematics to the real term, one has to decompose it into
contributions that are singular in only one kinematic region.  Since
$R_{g\bar{q},q}$ and $R_{qg,\bar{q}}$ are only singular in the $\splus$ and
$\sminus$ direction respectively, we assign their underlying Born to be the
same of the corresponding CS subtraction term. For $R_{q\bar{q},g}$, on the
other hand, we separate:
\begin{equation}
R_{q\bar{q},g}=R^\splus_{q\bar{q},g}+R^\sminus_{q\bar{q},g},\qquad\quad
R^\splusminus_{q\bar{q},g} = R_{q\bar{q},g}
\frac{C^\splusminus_{q\bar{q},g}}{C^\splus_{q\bar{q},g} +
  C^\sminus_{q\bar{q},g}}\;, 
\end{equation}
and assign to $R^\splusminus_{q\bar{q},g}$ the same underlying Born kinematics
of the corresponding CS counterterm $C^\splusminus_{q\bar{q},g}$.
The underlying Born flavour, on the other hand, is always $q\bar{q}$ in
the notation we have adopted.

\subsection{Generation of the Born variables}
The primary ingredient for a \POWHEG{} implementation is the $\bar{B}$
function, that is the inclusive cross section at fixed underlying Born
variables. In our case, it is given by
\begin{eqnarray}
\label{eq:bbarraqqbar}
\bar{B}&=&\sum_{q\bar{q}} \bar{B}_{q\bar{q}}, \\
\bar{B}_{q\bar{q}}&=&B_{q\bar{q}}({\bf \Phi}_2)+V_{q\bar{q}}({\bf \Phi}_2)
+ \sum_\splusminus \int \left[d {\Phi}_{\rm rad} \left\{
R_{q\bar{q},g}^\splusminus({\bf \Phi}_3)
-C^\splusminus_{q\bar{q},g}({\bf \Phi}_3)\right\}
\right]_\splusminus^{\bar{\bf \Phi}_2={\bf \Phi}_2}\nonumber
 \\ 
&+& \hspace{-1.7mm}\int \left[d {\Phi}_{\rm rad} \left\{
R_{g\bar{q},q}({\bf \Phi}_3)
-C_{g\bar{q},q}({\bf \Phi}_3)\right\}
\right]_\splus^{\bar{\bf \Phi}_2={\bf \Phi}_2}
+ \hspace{-1.7mm}\int\left[ d {\Phi}_{\rm rad}  \left\{
R_{qg,\bar{q}}({\bf \Phi}_3)
-C_{qg,\bar{q}}({\bf \Phi}_3)\right\}
\right]_\sminus^{\bar{\bf \Phi}_2={\bf \Phi}_2}
\nonumber \\
&+&\hspace{-1.7mm}
\int_{\bar{x}_\splus}^1 \!\! \frac{d z}{z} \left[
G_\splus^{q\bar{q},g}({\bf \Phi}_{2,\splus})
+ 
G_\splus^{g\bar{q},q}({\bf \Phi}_{2,\splus})
\right]^{\bar{\bf \Phi}_2={\bf \Phi}_2}+\hspace{-1.7mm}
 \int_{\bar{x}_\sminus}^1\!\! \frac{d z}{z} \left[
G_\sminus^{q\bar{q},g}({\bf \Phi}_{2,\sminus})
+
G_\sminus^{qg,\bar{q}}({\bf \Phi}_{2,\sminus})
\right]^{\bar{\bf \Phi}_2={\bf \Phi}_2}\nonumber\\
\end{eqnarray}
 The radiation variables $\Phi_{\rm rad}$
are parametrized in terms of three variables that span the unit cube, $X_{\rm
  rad}=\{X_{\rm rad}^{(1)}, X_{\rm rad}^{(2)},X_{\rm rad}^{(3)}\}$, while the
$z$ variable is parametrized in term of a single variable $X^{(1)}_{\rm rad}$
that ranges between 0 and 1. We then define the $\tilde{B}$ function
\begin{eqnarray}
\tilde{B}_{q\bar{q}}&=&B_{q\bar{q}}({\bf \Phi}_2)+V_{q\bar{q}}({\bf \Phi}_2)
+ \sum_\splusminus \left[\left|\frac{\partial{\Phi}_{\rm rad}}{\partial
    X_{\rm rad}} \right| \left\{
R_{q\bar{q},g}^\splusminus({\bf \Phi}_3)
-C^\splusminus_{q\bar{q},g}({\bf \Phi}_3) \right\}
\right]_\splusminus^{\bar{\bf \Phi}_2={\bf \Phi}_2}
\nonumber \\
&+& \left[ \left|\frac{\partial{\Phi}_{\rm rad}}{\partial X_{\rm rad}}
\right| \hspace{-1mm} \left\{
R_{g\bar{q},q}({\bf \Phi}_3)
-C_{g\bar{q},q}({\bf \Phi}_3) \right\}
\right]_\splus^{\bar{\bf \Phi}_2={\bf \Phi}_2}
\hspace{-3mm}+   \left[\left|\frac{\partial{\Phi}_{\rm rad}}{\partial X_{\rm
      rad}} 
\right| \hspace{-1mm}\left\{
R_{qg,\bar{q}}({\bf \Phi}_3)
-C_{qg,\bar{q}}({\bf \Phi}_3) \right\}
\right]_\sminus^{\bar{\bf \Phi}_2={\bf \Phi}_2}
\nonumber \\
&+&\lq \frac{1}{z} \frac{\partial z}{\partial X_{\rm rad}^{(1)}}  \lg
G_\splus^{q\bar{q},g}({\bf \Phi}_{2,\splus})
+ G_\splus^{g\bar{q},q}({\bf \Phi}_{2,\splus})\rg
\rq_\splus^{\bar{\bf \Phi}_2={\bf \Phi}_2} \nonumber\\
\label{eq:Btilde}
&+&
\lq \frac{1}{z}  \frac{\partial z}{\partial X_{\rm rad}^{(1)}}  \lg
G_\sminus^{q\bar{q},g}({\bf \Phi}_{2,\sminus})
+ G_\sminus^{qg,\bar{q}}({\bf \Phi}_{2,\sminus})
\rg \right]_\sminus^{\bar{\bf \Phi}_2={\bf \Phi}_2} \;,
\end{eqnarray}
so that defining $\tilde{B}=\sum_{q\bar{q}} \tilde{B}_{q\bar{q}}$, we have
\begin{equation}
\bar{B}=\int d^3 X_{\rm rad}\, \tilde{B}\,.
\end{equation}
In practice, the $\tilde{B}$ function is integrated numerically over all
${\bf \Phi}_2,X_{\rm rad}$ integration variables, using an integration program
that can generate the set of kinematic variables
${\bf \Phi}_2,X_{\rm rad}$, with a probability proportional to
$d{\bf \Phi}_2 \,d^3 X_{\rm rad}\, \tilde{B}$
in the $d{\bf \Phi}_2 \,d^3 X_{\rm rad}$ kinematic cell (see, for example,
refs.~\cite{Kawabata:1995th,Nason:2007vt}).  
Once the ${\bf \Phi}_2,X_{\rm rad}$ point is generated,
the flavour $q\bar{q}$ is chosen with a probability
proportional to the value of $\tilde{B}_{q\bar{q}}$ at that specific
${\bf \Phi}_2,X_{\rm rad}$ point.
At this stage, the radiation variables are disregarded, and only the
underlying Born ones are kept. This corresponds to integrate over the
radiation variables.

\subsection{Generation of the radiation variables}
Radiation kinematics is instead generated using the \POWHEG{} Sudakov form
factor
\begin{equation}
\Delta^{q\bar{q}}({\bf \Phi}_2,\pt)=
\prod_{\splusminus}\Delta^{q\bar{q}}_\splusminus\,,
\end{equation}
where
\begin{eqnarray}
\label{eq:suddef_ar1}
\Delta^{q\bar{q}}_\splus({\bf \Phi}_2,\pt) \!&=& \!
\exp\lg -\left[ \int d\Rad\,
\frac{R_{q\bar{q},g}^\splus({\bf \Phi}_3)
+R_{g\bar{q},q}({\bf \Phi}_3)}{B_{q\bar{q}}({\bf \Phi}_2)}
 \, \stepf\(\kt({\bf \Phi}_3)-\pt\) 
\right]_\splus^{\bar{\bf \Phi}_2={\bf \Phi}_2}\rg
\\
\label{eq:suddef_ar2}
\Delta^{q\bar{q}}_\sminus({\bf \Phi}_2,\pt) \!&=& \!
\exp\lg -\left[ \int d\Rad\,
\frac{R_{q\bar{q},g}^\sminus({\bf \Phi}_3)
+R_{qg,\bar{q}}({\bf \Phi}_3)}{B_{q\bar{q}}({\bf \Phi}_2)}
 \, \stepf\(\kt({\bf \Phi}_3)-\pt\) 
\right]_\sminus^{\bar{\bf \Phi}_2={\bf \Phi}_2}\rg\phantom{aaa}
\end{eqnarray}
The function $k_T({\bf \Phi}_3)$ measures the hardness of radiation in the
real event. It is required to be of the order of the transverse momentum of
the radiation in the collinear limit, and to become equal to it in the
soft-collinear limit. In principle, the choice of $k_T({\bf \Phi}_3)$ can
differ in the two singular regions ($\splus$ and $\sminus$) that we are
considering. The choice adopted in the Examples section of
ref.~\cite{Frixione:2007vw} had in fact this feature. We have found, however,
that for practical reasons\footnote{The choice discussed
in~\cite{Frixione:2007vw} is $k_T^2=M^2 (1-x) v$, and is such that $k_T^2$ is
always bound to be smaller than $M^2$.  Since the factorization and
renormalization scales are taken equal to $k_T$, for vector-boson production
at transverse momenta much larger than the vector-boson mass the coupling
does not properly decrease.}  it is better to adopt a different choice,
namely to take $k_T({\bf \Phi}_3)$ to coincide with that of
eqs.~(\ref{eq:ktdef}) and~(\ref{eq:ktsq}).

The generation of radiation is performed individually for
$\Delta^{q\bar{q}}_\splus$ and $\Delta^{q\bar{q}}_\sminus$, and the highest
generated $k_T$ is retained.  The upper bounding function for the application
of the veto method is chosen to be\footnote{This upper bounding function
  differs from the ones of eqs.~(7.163)--(7.166) in
  ref.~\cite{Frixione:2007vw}, but is in fact equivalent to the bound of
  eq.~(7.234) in the same reference, once the change of variables $\xi=1-x$,
  $y=(1-2v-x)/(1-x)$ is performed, and the different definitions of
  $d\Phi_{\rm rad}$ are properly taken into account.}
\begin{equation}\label{eq:uboundf}
\frac{R^\splus_{q\bar{q},g}+R_{g\bar{q},q}}{B_{q\bar{q}}}\le
\frac{16\pi^2}{M^2}\,
 N^\splus_{q\bar{q}}\,\frac{\alpha_s (k_T^2)}{2 v}\,
  \frac{x^2}{1 - x - v} \,,
\end{equation}
and the analogous one for the $\sminus$ direction.  The procedure used to
generate radiation events according to this upper bounding function is
described in Appendix~\ref{app:uboundfun}.

\subsection{Born zeros}
\label{sec:bornzeros}
In case the Born cross section vanishes in particular kinematics points, a
problem arises in the {\POWHEG} expression for the Sudakov form
factor~(\ref{eq:suddef_ar1}) and~(\ref{eq:suddef_ar2}). It happens, in fact,
that although $B$ vanishes, $\bar{B}$ may differ from zero.  Born kinematics
configurations with a vanishing Born cross section may thus be generated and,
at the stage of radiation generation, one would find very large ratios of the
real-emission cross section over the Born cross section. It would thus prove
difficult to find a reasonable upper bound for this ratio. If one tries to
neglect the problem, radiation events with a vanishing underlying Born
configuration would never be generated. We observe that, in the limit of
small hardness parameter, the real cross also exhibit the same vanishing
behaviour of the Born cross section. Loosely speaking, the problem arises
when the distance of the underlying Born configuration from the zero
configuration is smaller than the distance of the real emission cross section
from the singular (i.e.~zero hardness) configuration. In order to solve this
problem, in a completely general way, we further decompose the real cross
section contribution as (we use the notation of ref.~{\cite{Frixione:2007vw}})
\begin{equation}
  R^{\alpha_r} = R^{\alpha_r,s} + R^{\alpha_r,r},
\end{equation}
where
\begin{equation}
  R^{\alpha_r,s} = R^{\alpha_r} \, \frac{Z}{Z + H},\qquad\quad
  R^{\alpha_r,r} = R^{\alpha_r} \, \frac{H}{Z + H} .
  \label{eq:srpartition}
\end{equation}
The suffixes $s$ and $r$ stand for ``singular'' and ``regular'' respectively,
and $Z$ is a function of the kinematics that vanishes like the Born cross
section, evaluated at the underlying Born kinematics of the given term.  $H$
is the hardness of radiation and it must vanish for vanishing transverse
momentum of the radiation.  The simplest possible choice would be
\begin{equation}
  Z=  {\mathcal{B}}\, \frac{k^2_{T, \max}}{\mathcal{B}_{\max}},\quad\quad
  H=k_T^2\;,
\end{equation}
where $k_T$ is some definition of the transverse momentum of the radiation.
Notice now that $R^{\alpha_r,s}$ vanishes as fast as the Born term when its
underlying Born kinematics approaches the Born zero. It can thus be used in
the expression for the Sudakov form factor (eqs.~(\ref{eq:suddef_ar1})
and~(\ref{eq:suddef_ar2})) without problems.  The $R^{\alpha_r,r}$ is instead
non-vanishing, but, on the other hand, it does not have collinear or soft
singularities because of the $H$ factor, and thus it can be computed
directly, without any Sudakov form factor.  In the case of $W$ production,
the Born zero is associated to $\bar{\theta}_l=0$ if $q$ is an antiquark, and
$\bar{\theta}_l=\pi$ if it is a quark.  We choose then
\begin{equation}
Z=M^2\(1+s_q \,\cos \bar{\theta}_l\)^2,\qquad \quad    H=k_T^2\,,
\end{equation}
with $k_T^2$ given by formula~(\ref{eq:ktsq}) and the factor $s_q$ equals $1$
for quark, and $-1$ for antiquark. The angle $\bar\theta_l$ is chosen according
to the $\splus$ parametrization (for $R^\splus$) or the $\sminus$
parametrization (for $R^\sminus$) of the real-emission phase space.

In addition, all the $R^{\alpha_r}$ terms in eq.~(\ref{eq:bbarraqqbar}) are
replaced by the corresponding $R^{\alpha_r,s}$ and the 
$R^{\alpha_r,r}$ terms are generated in a way similar to what
was done for eq.~(\ref{eq:Btilde}). In other words one defines
\begin{eqnarray}
\tilde{B}^r=\sum_{q\bar{q}}\tilde{B}_{q\bar{q}}^r =\sum_{q\bar{q}}
&&\!\!\! \lg \left[\left|\frac{\partial{\Phi}_{\rm rad}}{\partial  X_{\rm rad}}
  \right|  
R_{q\bar{q},g}^{\splus,r}({\bf \Phi}_3) \right]_\splus^{\bar{\bf
     \Phi}_2={\bf \Phi}_2} +
 \left[\left|\frac{\partial{\Phi}_{\rm rad}}{\partial  X_{\rm rad}} \right| 
R_{q\bar{q},g}^{\sminus,r}({\bf \Phi}_3) \right]_\sminus^{\bar{\bf
     \Phi}_2={\bf \Phi}_2} \right. \nonumber\\
&&\!\!\!\! \left. +\left[ \left|\frac{\partial{\Phi}_{\rm rad}}{\partial X_{\rm
       rad}} \right| R^r_{g\bar{q},q}({\bf \Phi}_3) \right]_\splus^{\bar{\bf
     \Phi}_2={\bf \Phi}_2} +  \left[\left|\frac{\partial{\Phi}_{\rm
       rad}}{\partial X_{\rm rad}} \right| 
R^r_{qg,\bar{q}}({\bf \Phi}_3)\right]_\sminus^{\bar{\bf \Phi}_2={\bf \Phi}_2}
 \rg, \phantom{aaa}
\end{eqnarray}
and integrates over the whole  ${\bf \Phi}_2,X_{\rm rad}$ phase space
with the same method used for $\tilde{B}$.  In order to
generate an event, one chooses $\tilde{B}$ or $\tilde{B}^r$, with a
probability proportional to their respective total integral. In case
$\tilde{B}^r$ is chosen, one generates a kinematic configuration
according to it. This kinematic configuration is a full 3-body
configuration. The flavour $q\bar{q}$ is chosen with a probability
proportional the the value of $\tilde{B}_{q\bar{q}}^r$ for the particular
kinematic point that has been generated, and the event is sent to the output.
In case $\tilde{B}$ is chosen, a kinematic configuration and an underlying
Born flavour is chosen in the same way.

\section{Results}
\label{sec:results}
The \MCatNLO{} program provides an implementation of vector-boson production
at the NLO level in a shower Monte Carlo framework.  It should therefore be
comparable to our calculation, and we thus begin by comparing \MCatNLO{} and
\POWHEG{} distributions. In this comparison, the \POWHEG{} code is interfaced
to \HERWIG{} \cite{Corcella:2000bw,Corcella:2002jc},
in order to minimize differences due to the subsequent shower in
the two approaches.  We choose as our default parton-density functions the
CTEQ6M~\cite{Pumplin:2002vw} package, and the corresponding value of
$\LambdaQCD$. The factorization and renormalization scales are taken equal to
$M_V^2+(p_T^V)^2$ in the calculation of the $\bar{B}$ function, where $V=W$
or $Z$.  In the generation of radiation, the factorization and
renormalization scales are taken equal to the transverse momentum of the
vector boson $V$. We also account properly for the heavy-flavour thresholds,
when the transverse momentum of the vector boson approaches the bottom and
charm quark threshold.  That is to say, when the renormalization scale
crosses a heavy-flavour mass threshold, the QCD evolution of the running
coupling is accordingly changed to the new number of active flavours.
The other relevant parameters for our calculation are
\begin{center}
\begin{tabular}{|c|c|c|c|c|c|}
\hline
$M_z$ (GeV) & $\Gamma_Z$ (GeV) & $M_W$ (GeV) & $\Gamma_W$ (GeV) & 
$\sin^2\theta_W^{\rm eff}$&$\alpha^{-1}_{\rm em}(M_Z)$  \\ \hline
91.188 & 2.49 & 80.419 & 2.124 & 0.23113 & 127.934 \\
\hline
\end{tabular}\;.
\end{center}
The above values of masses and widths are used in eqs.~(\ref{eq:wprop})
and~(\ref{eq:Zgamprop}). The $W$ and $Z$ couplings are given by 
\begin{equation}
g=\frac{e}{\sin\theta_W^{\rm eff}}\,, \quad 
g_{l/q}=\frac{e}{\sin\theta_W^{\rm eff}\cos\theta_W^{\rm eff}} 
\left[T_3^{(l/q)}-q_{l/q}\sin^2\theta_W^{\rm eff}\right],
\quad e=\sqrt{4\pi \alpha_{\rm em}(M_Z)}\,,
\end{equation}
where $l/q$ denotes the given left or right component of a lepton or a
quark. For $W$ production we used the following absolute values
for the CKM matrix elements
\begin{center}
\begin{tabular}{|c|c|c|c|c|c|c|c|c|}
\hline
$ud$   & $us$   & $ub$   & $cd$   & $cs$   & $cb$  & $td$  & $ts$   & $tb$   \\ \hline
0.9748 & 0.2225 & 0.0036 & 0.2225 & 0.9740 & 0.041 & 0.009 & 0.0405 & 0.9992 \\
\hline
\end{tabular}\;.
\end{center}
In all figures shown in the following we do not impose any acceptance cut.
\subsection{$\boldsymbol Z$ production at the Tevatron}
\begin{figure}[ht]
\begin{center}
\epsfig{file=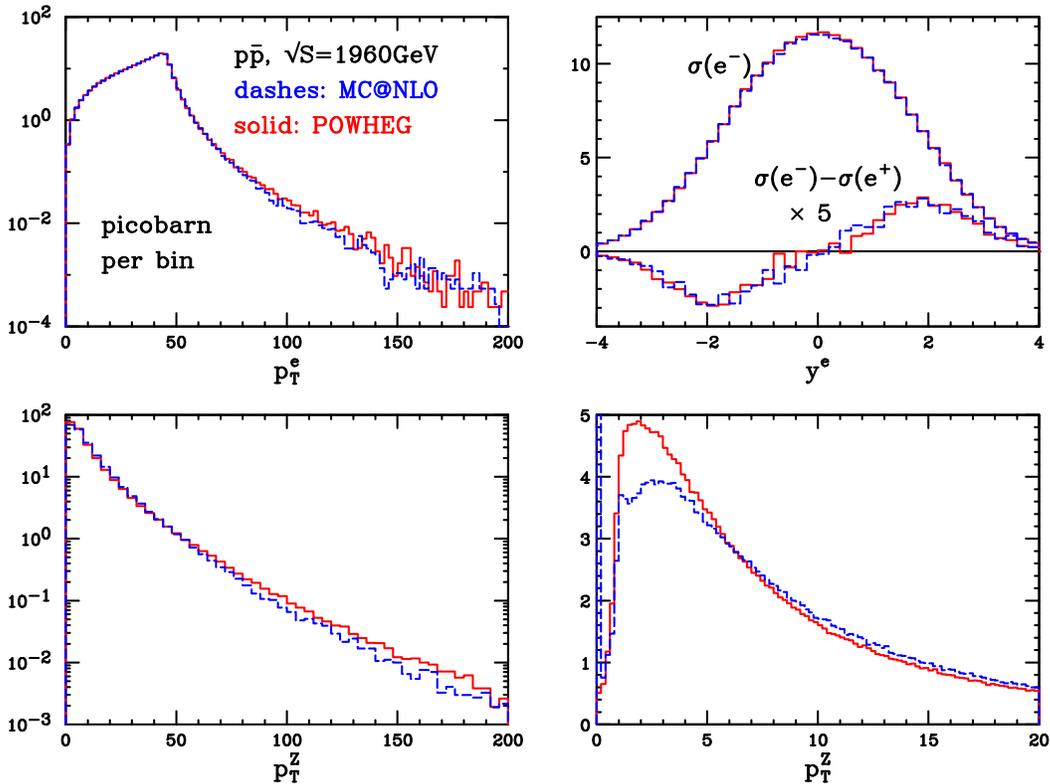,width=\figwidth}
\end{center}
\captskip
\caption{\label{fig:cmp1} Comparison between \POWHEG{} and \MCatNLO{} results
  for the transverse momentum and rapidity of the lepton coming from the
  decay of the $Z$ boson, and for the transverse momentum of the $Z$, as
  reconstructed from its decay products. The lepton-rapidity asymmetry is also
  shown. Plots done for the Tevatron $p\bar{p}$ collider.}
\end{figure}

\begin{figure}[ht]
\begin{center}
\epsfig{file=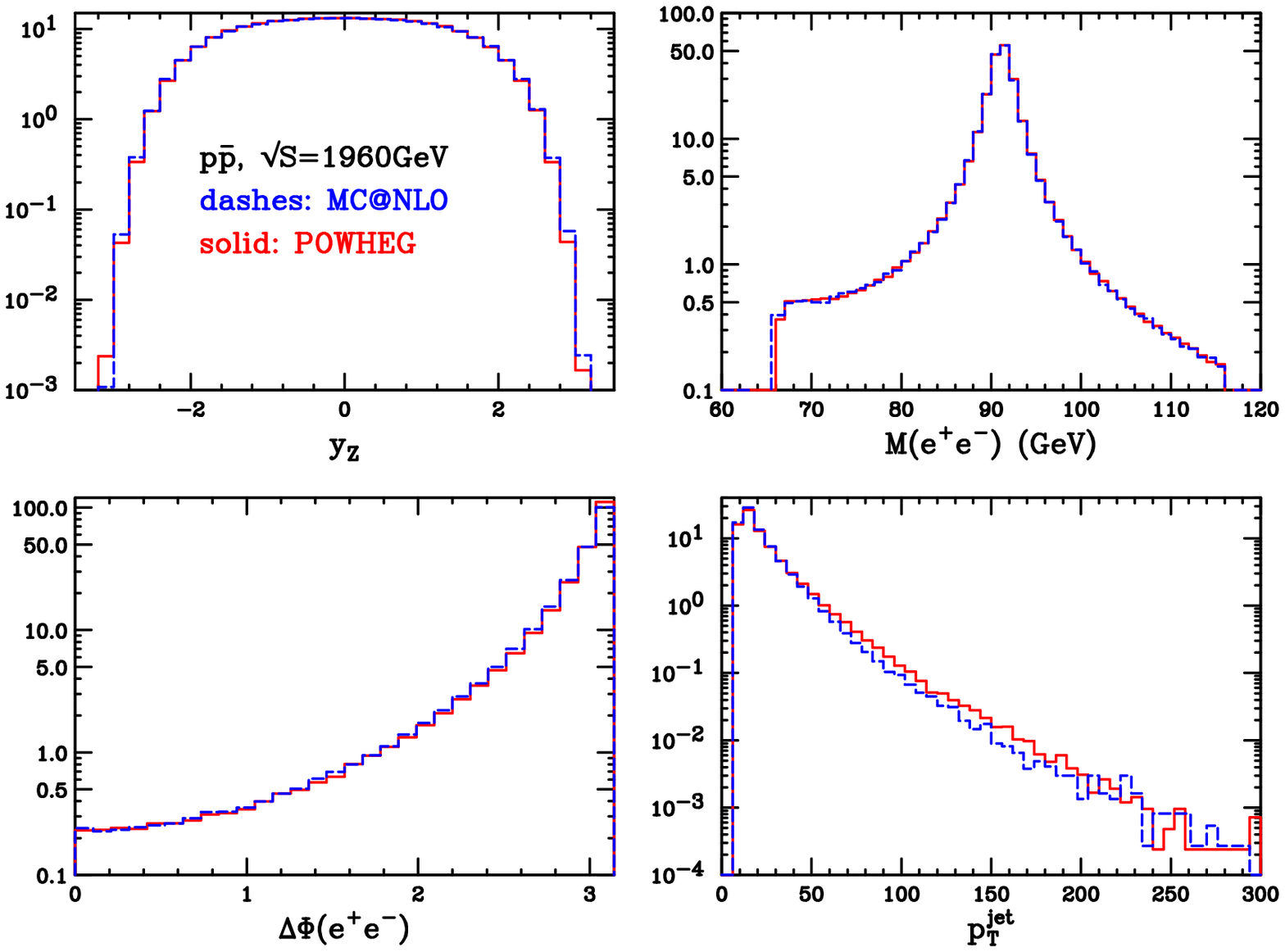,width=\figwidth}
\end{center}
\captskip
\caption{\label{fig:cmp2} Comparison between \POWHEG{} and \MCatNLO{} for the
  reconstructed $Z$ rapidity, its invariant mass, the lepton-pair azimuthal
  distance and the transverse momentum of the reconstructed jet, above a
  10~GeV minimum value.} 
\end{figure}

In fig.~\ref{fig:cmp1} we show a comparison of the lepton transverse momentum
and rapidity, and of the transverse momentum of the reconstructed
lepton-antilepton pair at the Tevatron.
We notice a larger cross section in \POWHEG{}, when the $Z$ transverse
momentum becomes large. This is not unexpected, since for large momenta the
\POWHEG{} result is larger than the standard NLO result by a factor
$\bar{B}/B$ (this feature has also some impact upon the transverse-momentum
distribution of the lepton).  Once this fact is accounted for, the
transverse-momentum distribution of the $Z$ is in fair agreement, although we
find 
observable shape differences at low transverse momenta. We also notice a peak
at $p_T=0$ in the \MCatNLO{} distribution, that is not present in the
\POWHEG{} result. We expect this distribution to be affected by low
transverse-momentum power-suppressed effects. In fact, the peak at zero
transverse momentum in \MCatNLO{} disappears if the primordial transverse
momentum of the partons (the {\tt PTRMS} variable in \HERWIG{}) is set to a
non-zero value.  In fig.~\ref{fig:cmp2} we compare the rapidity distribution
of the reconstructed $Z$, its invariant mass, the azimuthal distance of the
$e^+e^-$ pair coming from $Z$ decays, and the transverse momentum of the
radiated jet at the Tevatron.
The jet is defined using the \SISCONE{} algorithm~\cite{Salam:2007xv} as
implemented in the {\tt FASTJET} package~\cite{Cacciari:2005hq}, using
$R=0.7$. We find again fair agreement.

\begin{figure}[ht]
\begin{center}
\epsfig{file=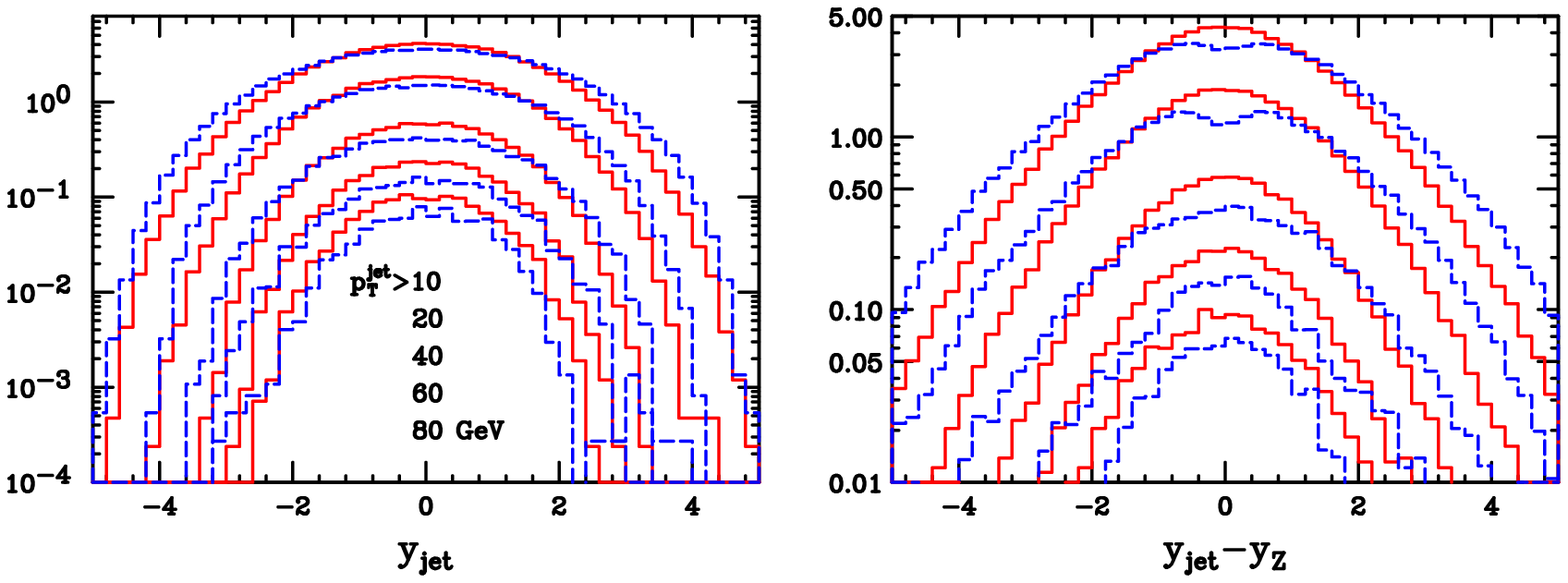,width=\figwidth}
\end{center}
\captskip
\caption{\label{fig:cmp3} Rapidity distribution of the hardest jet with
  different transverse-momentum cuts, and the rapidity distance between the
  hardest jet and the reconstructed $Z$ boson.}
\end{figure}

In ref.~\cite{Mangano:2006rw}, a discrepancy was found in the rapidity
distribution of the hardest radiated jet as computed in \MCatNLO{} and
\ALPGEN, for the case of top pair production at the Tevatron.  The \MCatNLO{}
calculation shows there a dip at zero rapidity, not present in \ALPGEN.  In
fact, the \POWHEG{} calculation of this quantity does not display any dip.
We thus examine the transverse momentum of the radiated jet in this case.
Furthermore, we also plot the rapidity difference between the $Z$ and the
hardest radiated jet. The results are displayed in fig.~\ref{fig:cmp3}.
We have chosen different cuts for the minimum transverse momentum of the
radiated jet, i.e. 10, 20, 40, 60 and 80~GeV.  We observe noticeable
differences in the rapidity distribution of the hardest jet in the two
approaches. The \MCatNLO{} result displays a dip at zero $y_{\rm jet}-y_Z$.

\subsection{$\boldsymbol Z$ production at the LHC}
\begin{figure}[ht]
\begin{center}
\epsfig{file=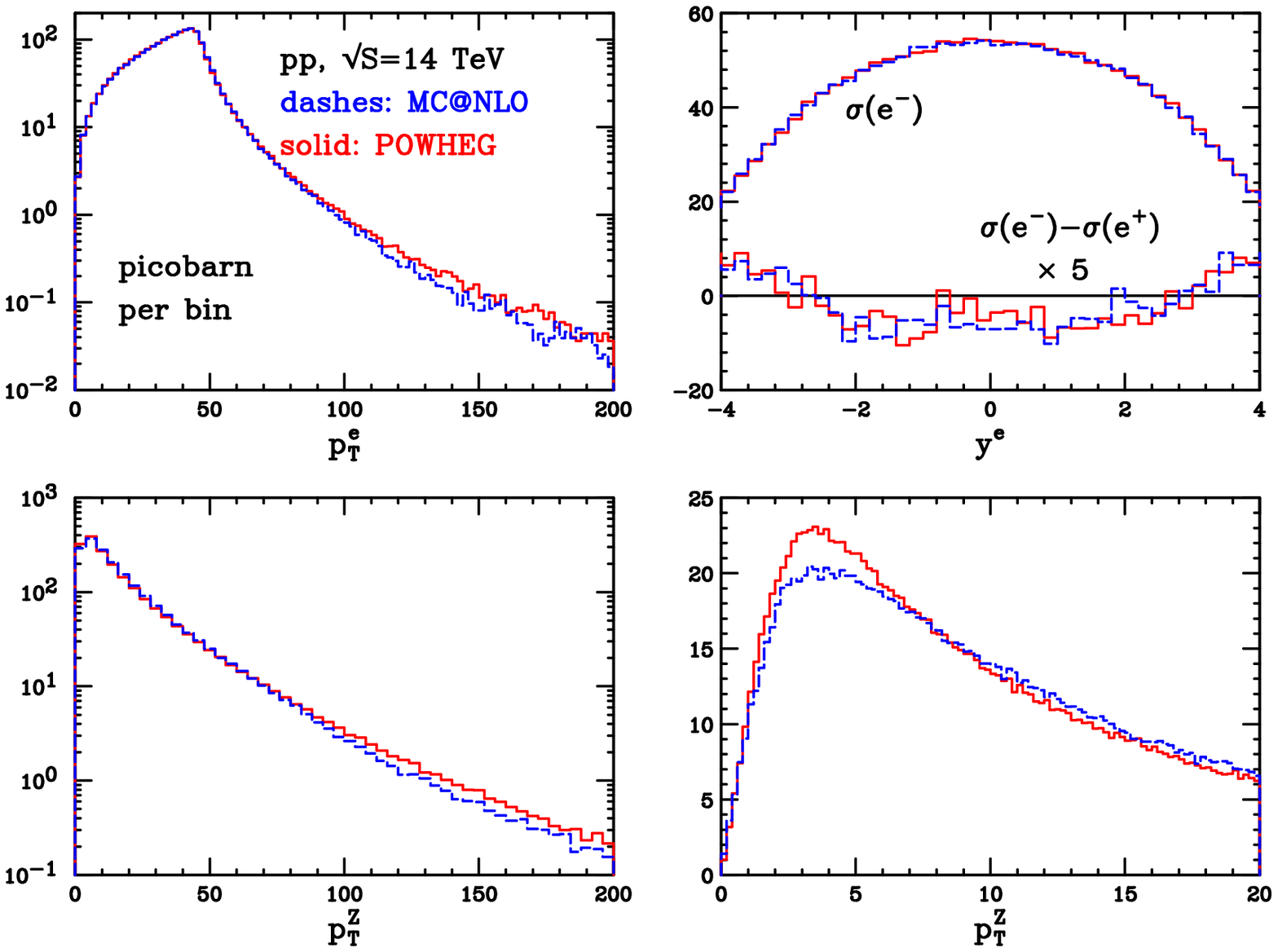,width=\figwidth}
\end{center}
\captskip
\caption{\label{fig:cmp1-lhc}
Same as fig.~\ref{fig:cmp1} 
for the LHC at 14~TeV.}
\end{figure}

\begin{figure}[ht]
\begin{center}
\epsfig{file=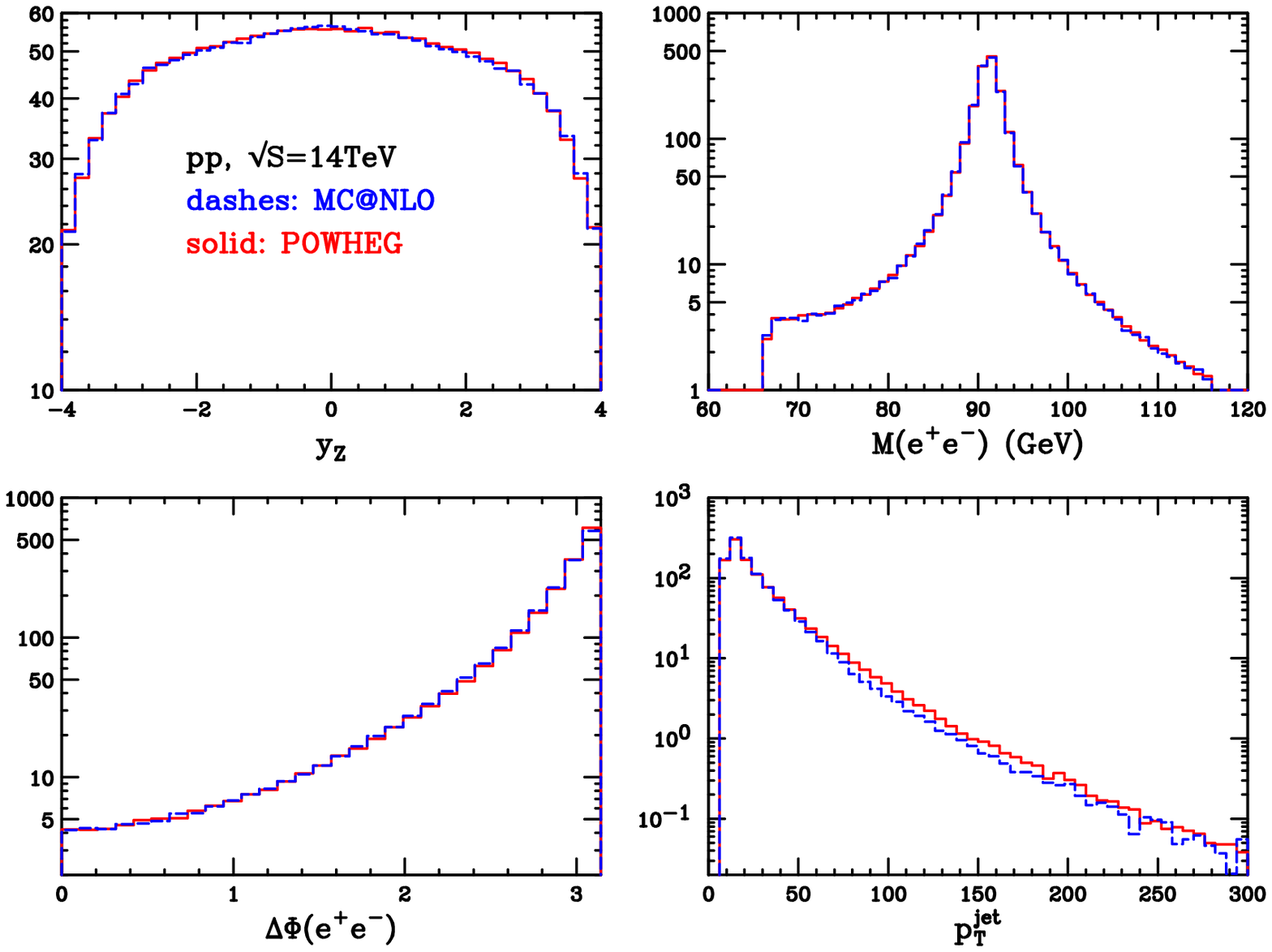,width=\figwidth}
\end{center}
\captskip
\caption{\label{fig:cmp2-lhc}
Same as fig.~\ref{fig:cmp2} 
for the LHC at 14~TeV.}
\end{figure}

\begin{figure}[ht]
\begin{center}
\epsfig{file=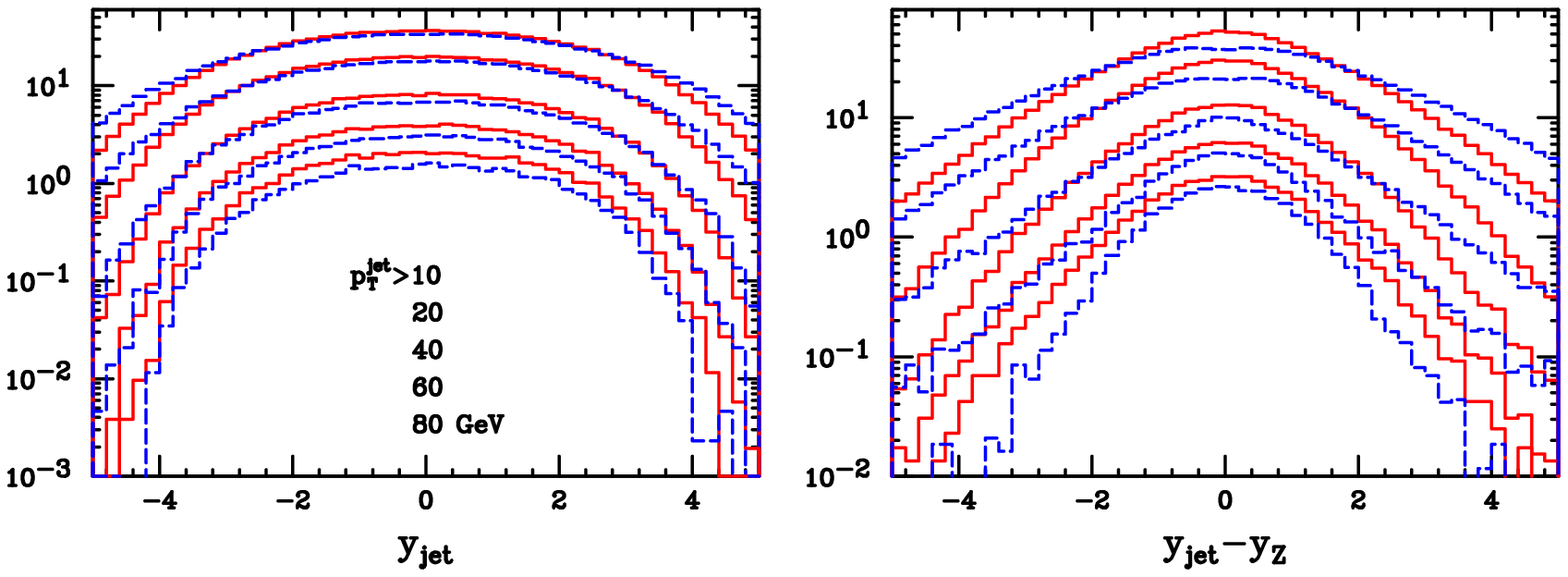,width=\figwidth}
\end{center}
\captskip
\caption{\label{fig:cmp3-lhc}
Same as fig.~\ref{fig:cmp3}
at the LHC at 14~TeV.}
\end{figure}

Similar results are reported for the LHC in fig.~\ref{fig:cmp1-lhc}
through~\ref{fig:cmp3-lhc}.
We notice less pronounced differences (with respect to the Tevatron case) in the $p_T$
spectrum of the $Z$ boson. The discrepancy in the $y_{\rm jet}$ distribution
is still evident, although the dip is barely noticeable in this case.
\begin{figure}[ht] %
\begin{center}
\epsfig{file=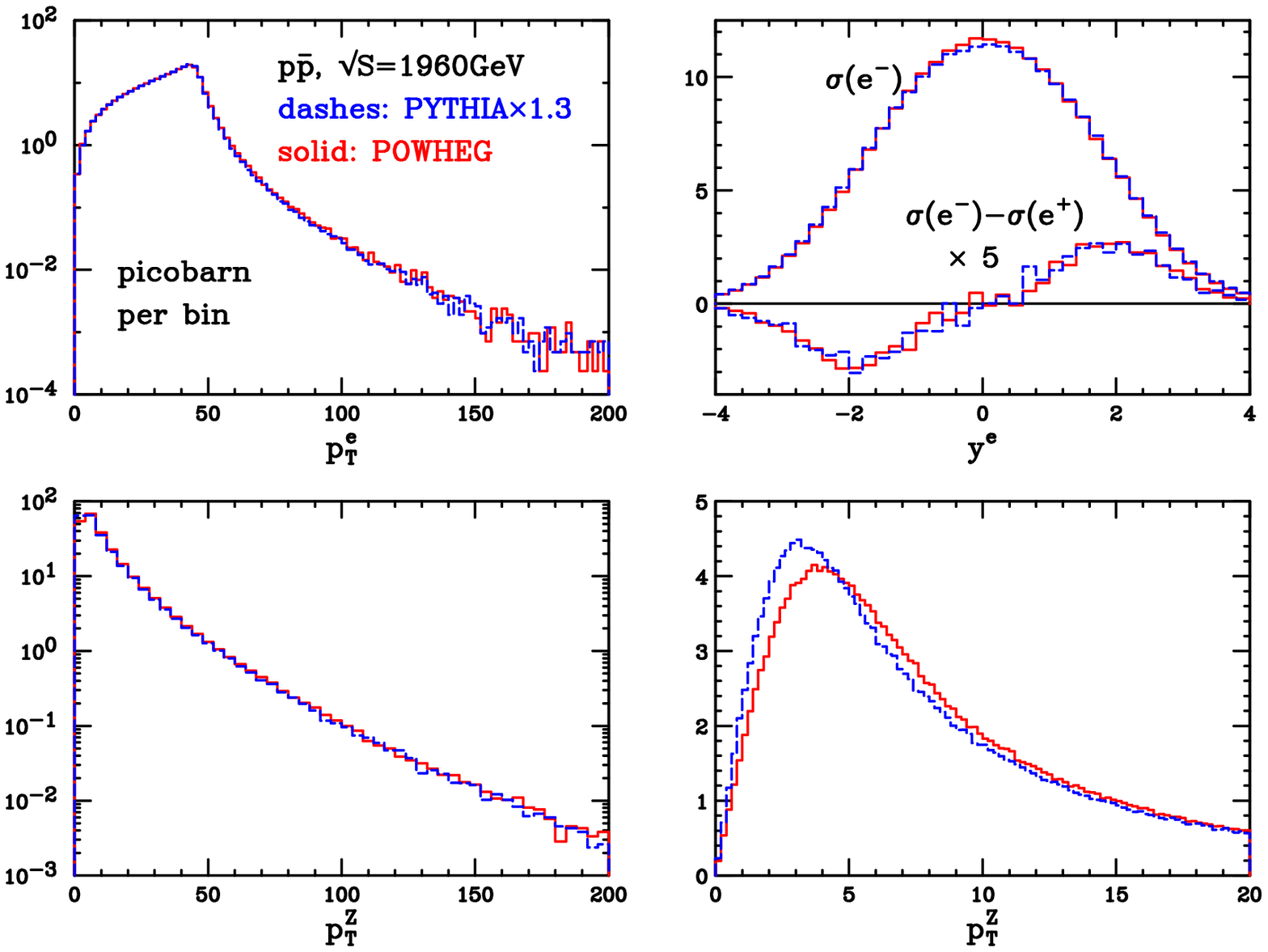,width=\figwidth}
\end{center}
\captskip
\caption{\label{fig:cmp1-py}
Same as fig.~\ref{fig:cmp1}
for a \PYTHIA{} and \POWHEG{} comparison at the Tevatron.}
\end{figure} %
\begin{figure}[ht] %
\begin{center}
\epsfig{file=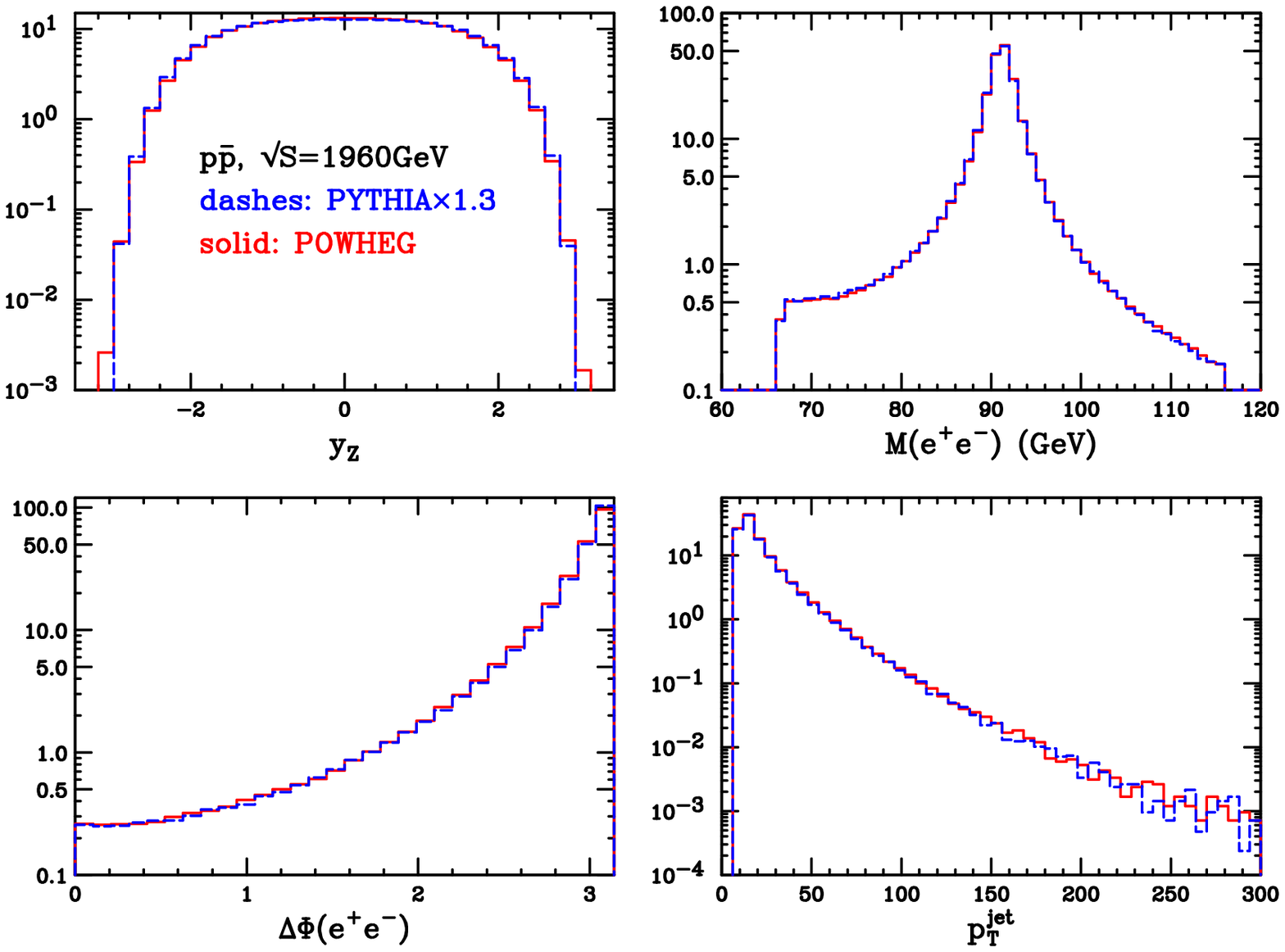,width=\figwidth}
\end{center}
\captskip
\caption{\label{fig:cmp2-py}
Same as fig.~\ref{fig:cmp2}
for a \PYTHIA{} and \POWHEG{} comparison at the Tevatron.}
\end{figure} %
\begin{figure}[ht] %
\begin{center}
\epsfig{file=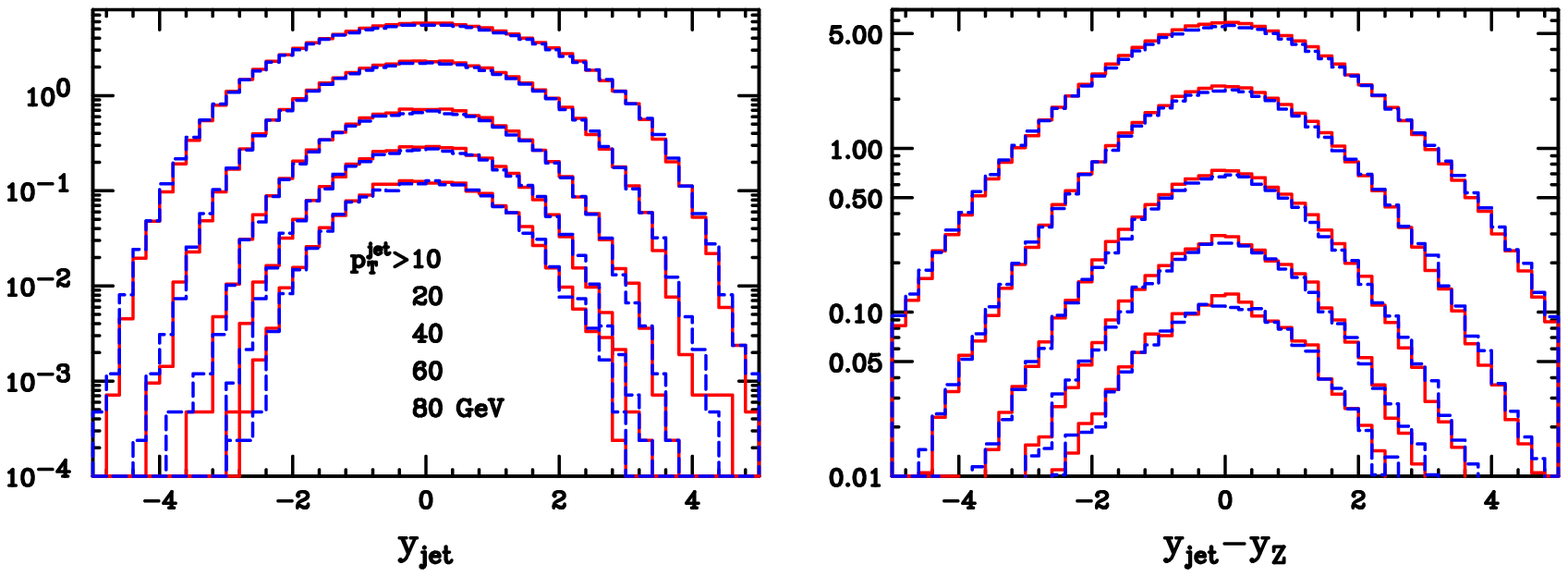,width=\figwidth}
\end{center}
\captskip
\caption{\label{fig:cmp3-py} %
Same as fig.~\ref{fig:cmp3}
for a \PYTHIA{} and \POWHEG{} comparison at the Tevatron.}
\end{figure}%
\begin{figure}[ht] %
\begin{center}
\epsfig{file=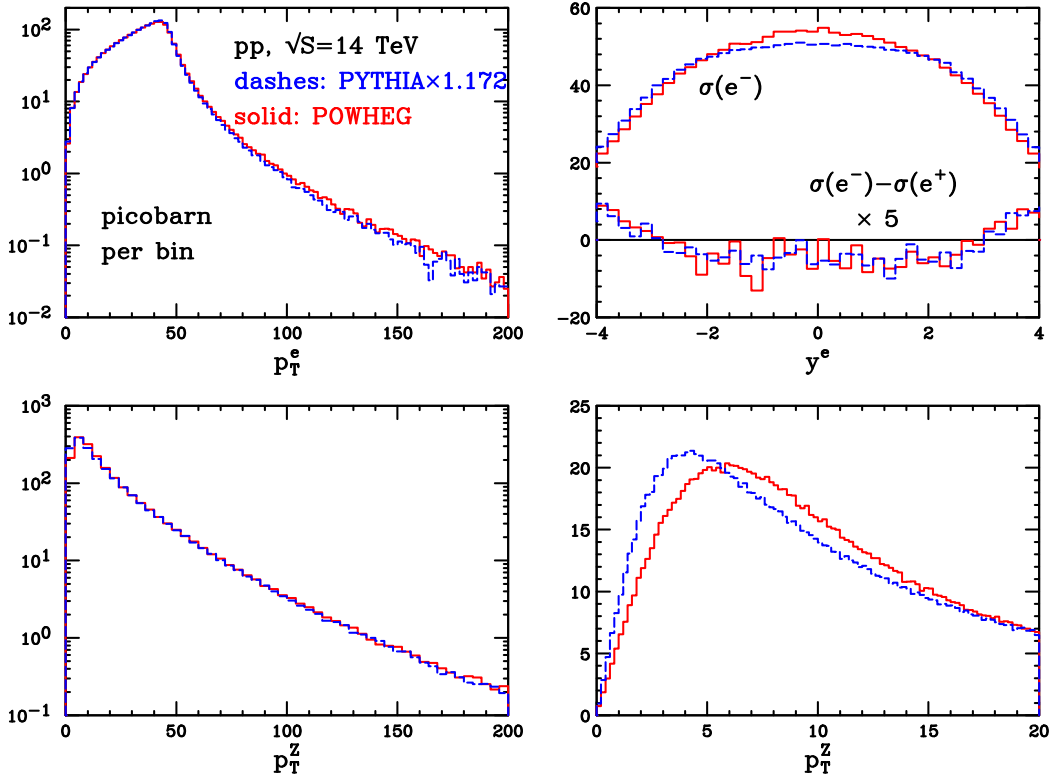,width=\figwidth}
\end{center}
\captskip
\caption{\label{fig:cmp1-lhc-py} %
Same as fig.~\ref{fig:cmp1}
for a \PYTHIA{} and \POWHEG{} comparison at the LHC.}
\end{figure}%
\begin{figure}[ht] %
\begin{center}
\epsfig{file=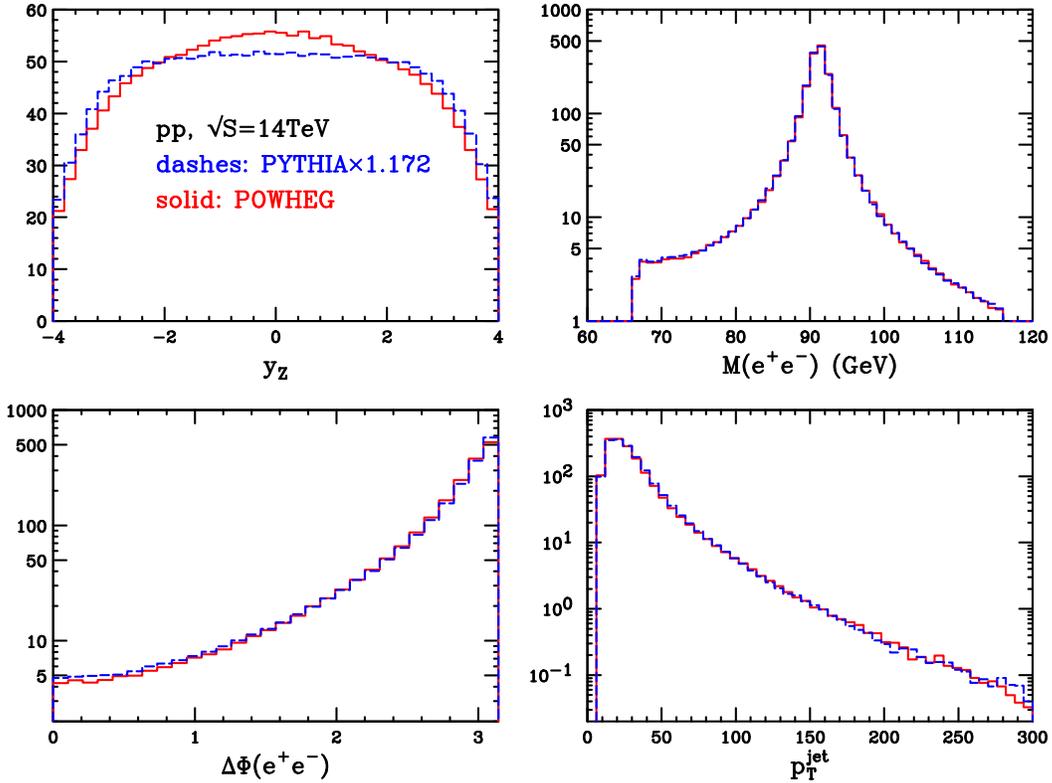,width=\figwidth}
\end{center}
\captskip
\caption{\label{fig:cmp2-lhc-py} %
Same as fig.~\ref{fig:cmp2}
for a \PYTHIA{} and \POWHEG{} comparison at the LHC.}
\end{figure} %
\begin{figure}[ht] %
\begin{center}
\epsfig{file=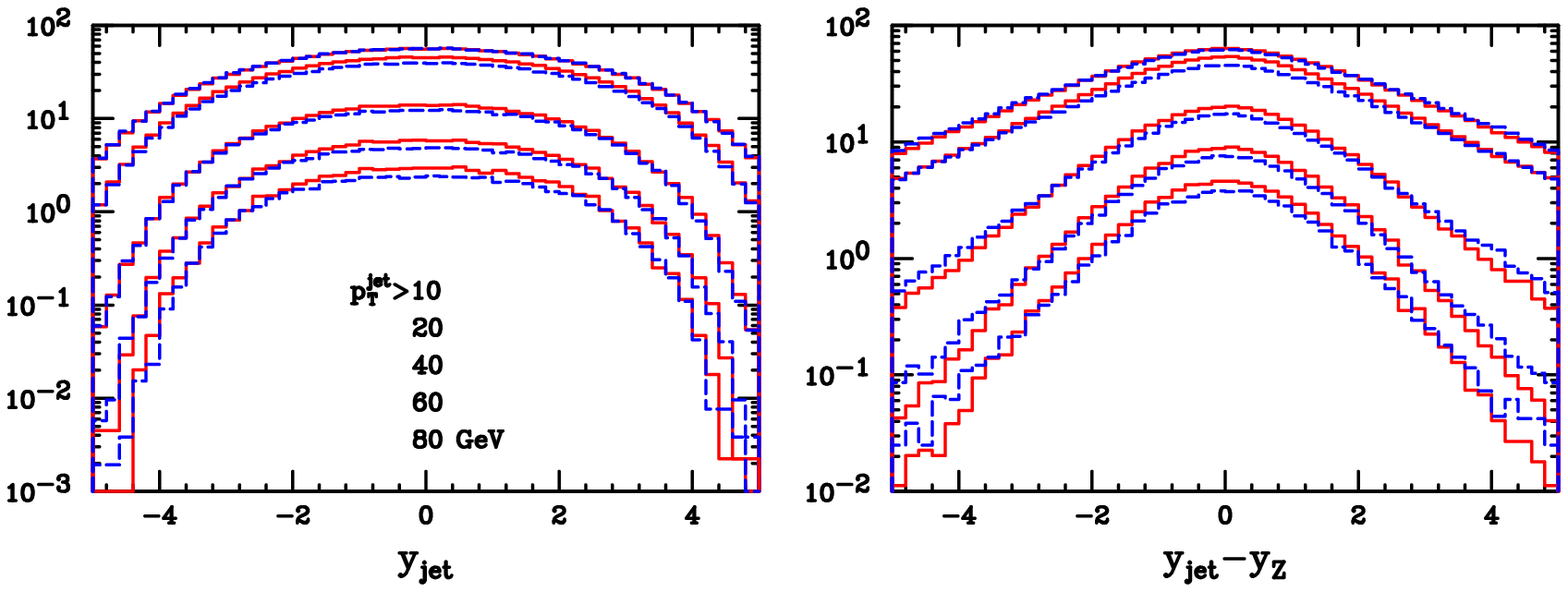,width=\figwidth}
\end{center}
\captskip
\caption{\label{fig:cmp3-lhc-py}  %
Same as fig.~\ref{fig:cmp3}
for a \PYTHIA{} and \POWHEG{} comparison at the LHC.}
\end{figure} %

The same set of plots are also shown for a \PYTHIA{}-\POWHEG{} comparison in
fig.~\ref{fig:cmp1-py} through~\ref{fig:cmp1-lhc-py}. In this case the
\POWHEG{} code was interfaced with \PYTHIA{}. Photon radiation from
final-state leptons was switched off ({\tt MSTJ(41)=3}), in order to simplify
the analysis. Furthermore, the new transverse-momentum ordered shower was
used (i.e.\ the {\tt PYEVNW} routine), since transverse-momentum ordering
should be more appropriate in conjunction with \POWHEG{}. In the plots, the
\PYTHIA{} output is normalized to the \POWHEG{} total cross section.  From
fig.~\ref{fig:cmp1-py} through~\ref{fig:cmp2-py}, we can see a remarkable
agreement between the two calculations for the Tevatron results, the only
visible discrepancy being given by the transverse-momentum distribution of
the $Z$ boson at small transverse momenta.  We also notice that, unlike the
case of the \MCatNLO{}-\POWHEG{} comparison, the transverse-momentum
distribution of the $Z$ is slightly harder in \PYTHIA{} than in
\POWHEG{}. The rapidity distributions of the hardest jet are also in
remarkable agreement.

\begin{figure}[ht] %
\begin{center}
\epsfig{file=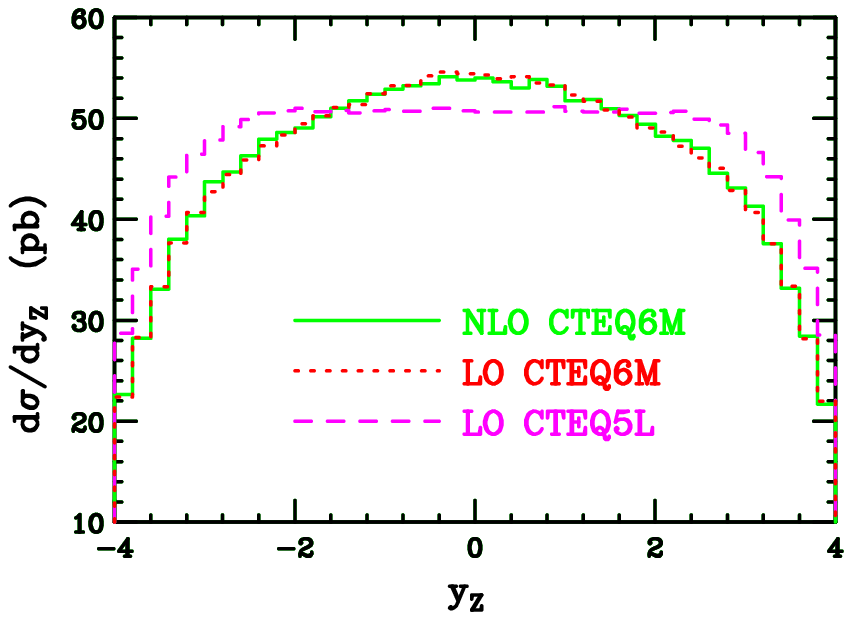,width=0.5\figwidth}
\end{center}
\captskip
\caption{\label{fig:yzLONLO} 
Rapidity distribution for the $Z$ boson, computed at fixed order at LO and
NLO.  For the LO result, both the CTEQ6M and the CTEQ5L parton-density set
were used.  The plots are normalized to the NLO total cross section.}
\end{figure}

In fig.~\ref{fig:cmp1-lhc-py} through~\ref{fig:cmp2-lhc-py}, we carry out the
same comparison in the LHC case. We notice here few important differences in
the rapidity distribution of the $Z$ boson, and, probably related to that, of
the electron, the \PYTHIA{} distribution being flatter in the central
region. Both \MCatNLO{} and \POWHEG{} do not show this feature.  As already
pointed out in ref.~\cite{Frixione:2007vw}, the generation of vector bosons
in \PYTHIA{} is not very different from the \POWHEG{} generation. Radiation
is generated with a very similar
method~\cite{Bengtsson:1986hr,Sjostrand:2006su}.  There are however
differences.  In \PYTHIA{} the Born inclusive cross section is used rather
than our $\bar{B}$ function.  Furthermore, our choice of scales is
constrained by the requirement of next-to-leading logarithmic accuracy in the
Sudakov form factor. The discrepancy in the transverse-momentum distribution
of the $Z$ may be due to different requirements for the choice of the scale
in the generation of radiation in the two algorithms. The discrepancy in the
rapidity distribution may be due to the lack of NLO corrections in \PYTHIA{},
i.e.\ to the use of the Born cross section (rather than the $\bar{B}$
function) and LO parton densities.  In fact, in fig.~3 of
ref.~\cite{Anastasiou:2003ds}, a comparison in the rapidity distribution of
the $Z$ at LO, NLO and NNLO, is shown for the LHC. One can notice from that
figure that there is a difference in the LO and NLO shape of the
distribution, the former being flatter.  In order to elucidate this point, we
show in fig.~\ref{fig:yzLONLO} the rapidity distribution of the $Z$ boson
computed at fixed order in QCD, at LO and NLO. With the LO calculation, we
also show the result obtained using the same LO parton-distribution function
(pdf) set used in \PYTHIA, that is CTEQ5L.
The figure leads to the conclusion  that the use of the LO parton-density set
CTEQ5L is  the primary cause of this  shape difference. We find,  in fact, no
difference in shape between the LO and NLO result if the same pdf set is used
instead.  We  thus  conclude that  also  the  effect  observed in  fig.~3  of
ref.~\cite{Anastasiou:2003ds} is  due to the  use of a LO  parton-density set
together  with the  LO  result.\footnote{Some  authors do  prefer  to use  LO
parton-density functions in LO  calculations, although, in our opinion, there
are no compelling reasons to do so.}

\begin{figure}[ht] %
\begin{center}
\epsfig{file=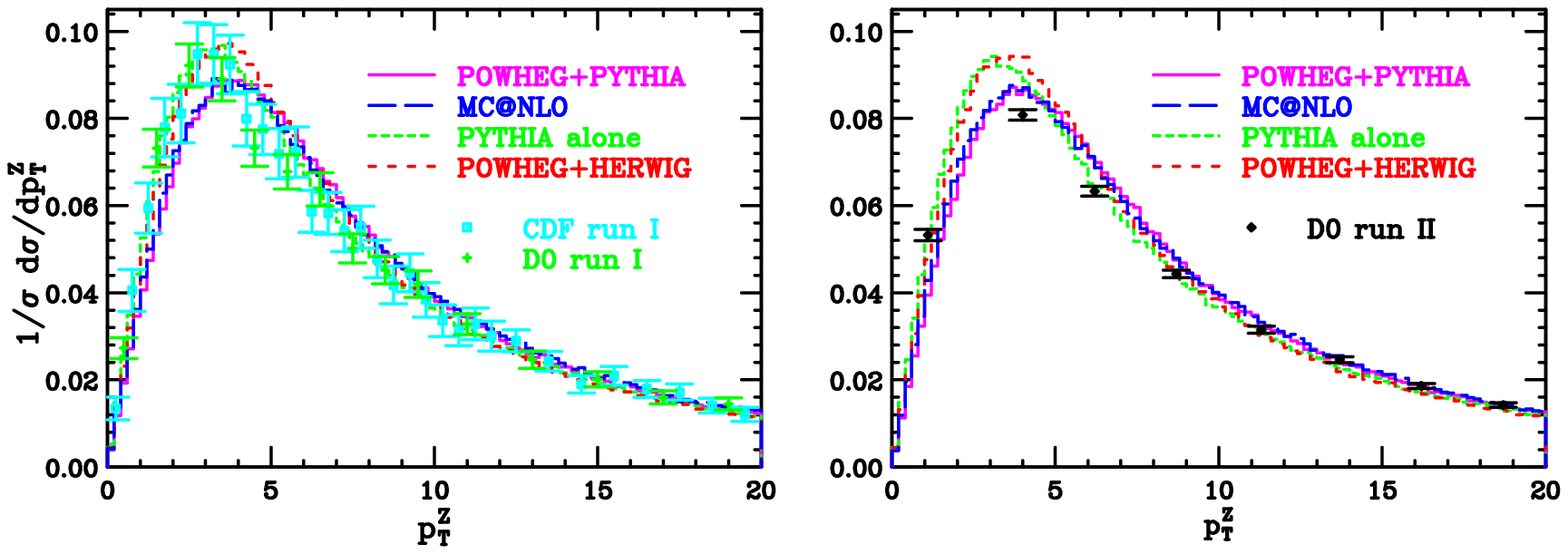,width=\figwidth}
\end{center}
\captskip
\caption{\label{fig:z-pt-comp} 
Comparison of transverse-momentum distributions of the $Z$ bosons with data
from the Tevatron.}
\end{figure}

The predictions for the transverse-momentum distribution of the $Z$ boson are
summarized in fig.~\ref{fig:z-pt-comp}, in comparison with data from
ref.~\cite{Abazov:2007nt}, at $\sqrt{S}=1960$~GeV and from
refs.~\cite{Affolder:1999jh,Abbott:1999wk,Abbott:1999yd} at
$\sqrt{S}=1800$~GeV.
The \POWHEG{}+\HERWIG{} and the \MCatNLO{} output are obtained with an
intrinsic transverse momentum of the incoming partons equal to $2.5$~GeV
(\HERWIG{}'s {\tt PTRMS} parameter).  Both data and predictions are
normalized to 1. The difference in the shape of the distributions at 1960 and
1800~GeV are only minimal. We see that \POWHEG{} with \PYTHIA{} is in
remarkable agreement with the \MCatNLO{} result. On the other hand,
standalone \PYTHIA{} is closer to the output of \POWHEG{} with \HERWIG{}. In
all cases, the agreement with data is not optimal.  It is thus clear that
this distributions is sensitive to long distance effects like hadronization
and transverse-momentum smearing, and good agreement with data may only
achieved by suitable tuning of the non-perturbative parameters of the shower
Monte Carlo.

\clearpage

\subsection{Hardest-jet rapidity distribution}
\begin{figure}[ht] %
\begin{center}
\epsfig{file=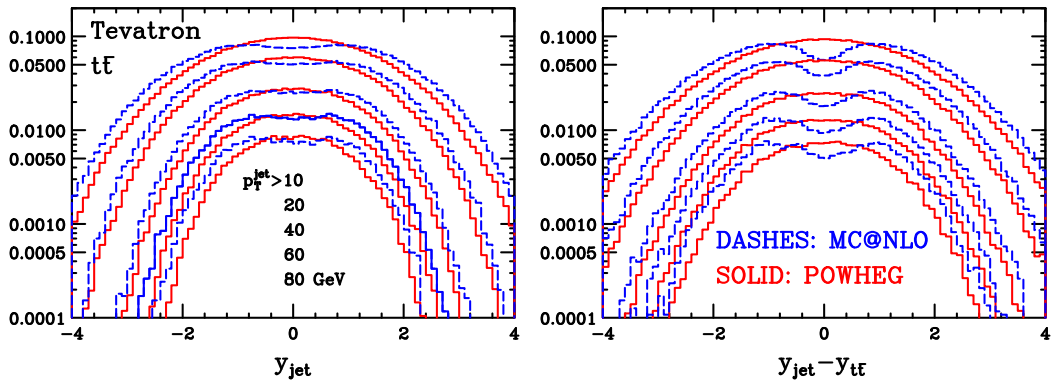,width=\figwidth}
\end{center}
\captskip
\caption{\label{fig:ttjets}  %
Rapidity distribution of the hardest jet and of the rapidity difference
between the hardest jet and the $t\bar{t}$ system at Tevatron energies.}
\end{figure}

\begin{figure}[ht] %
\begin{center}
\epsfig{file=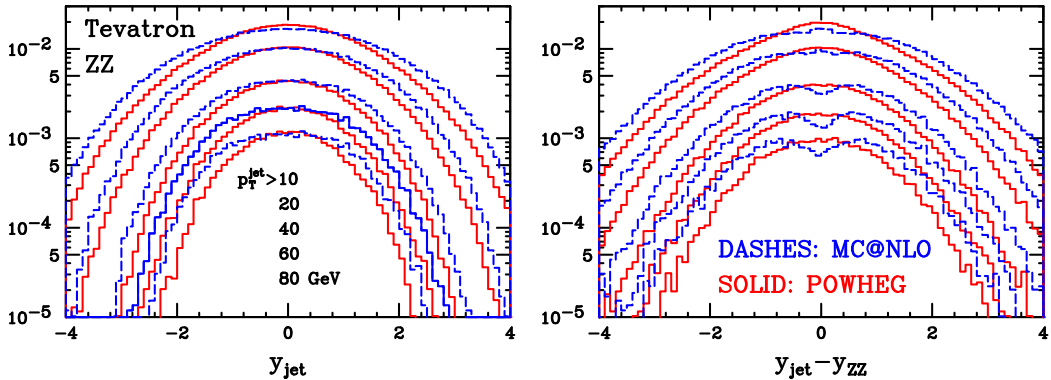,width=\figwidth}
\end{center}
\captskip
\caption{\label{fig:zzjets}  
Rapidity distribution of the hardest jet and of the rapidity difference
between the hardest jet and the $ZZ$ system at Tevatron energies.}
\end{figure}

The discrepancy of \POWHEG{} and \MCatNLO{} in the rapidity distribution of
the hardest jet deserves further discussion.\footnote{The distribution in the
  pseudorapidity difference of the hardest jet with respect to the vector
  boson was considered in ref.~\cite{Alwall:2007fs}, in the context of a
  comparison of several matrix-element programs. Although noticeable
  differences are found among the generators considered there, none of them
  exhibit a dip at zero pseudorapidity.}  In ref.~\cite{Mangano:2006rw}, only
the rapidity distribution of the hardest jet in $t\bar{t}$ production was
considered, and a dip was found there, in the case of top-pair production at
Tevatron energies.  In the present case we found no dip in 
the rapidity distribution of the hardest jet in $V$ production (see
fig.~\ref{fig:cmp3}). 
We found instead a dip in the distribution in the rapidity
difference between the jet and the vector boson. It is reasonable to assume
that a dip in the rapidity distribution of the jet may be inherited from the
dip in the rapidity difference, if the kinematics production regime is
forced to be central, like in the case of top-pair production at the
Tevatron. We 
thus also reconsider $Z$ pair production and $t\bar{t}$ production at the
Tevatron, and compare \POWHEG{} and \MCatNLO{} results for the rapidity
distribution of the hardest jet, and for the distribution in the rapidity
difference. The results are shown in figs.~\ref{fig:ttjets}
and~\ref{fig:zzjets}.
From fig.~\ref{fig:ttjets} we see that the dip present in the $y_{\rm
  jet}-y_{t\bar{t}}$ distribution is even deeper than the dip observed in the
$y_{\rm jet}$ distribution.  Furthermore, in fig.~\ref{fig:zzjets}, we see no
particular features in the $y_{\rm jet}$ distribution. The $y_{\rm
  jet}-y_{ZZ}$ distribution displays instead a tiny tower and a dip,
depending upon the transverse-momentum cut on the jet.  A deeper study of
these features was performed in ref.~\cite{Mangano:2006rw}, for $t\bar{t}$
production. It was shown there that the \HERWIG{} Monte Carlo displays an
even stronger dip than \MCatNLO{}. The \MCatNLO{} generator provides more
events that partially fill the dip, thus correcting the NLO inaccuracies of
the shower Monte Carlo. It is presumably a NNLO (next-to-next-to leading)
mismatch between the twos that generates these features. On the other hand,
the \POWHEG{} program, as well as matrix-element generators, generate
themselves the full NLO result, and thus are not sensitive to this feature of
\HERWIG{}.  We also stress that these features do not mean that \HERWIG{} is
inaccurate at the LO level, or that \MCatNLO{} is inaccurate at the NLO. A
shower Monte Carlo is accurate in the radiation of the hardest jet only in
the collinear regions. Furthermore, the dip in the \MCatNLO{} result is
compatible with an effect beyond NLO.

\subsection{${\boldsymbol W}$ production at the Tevatron and LHC}
\begin{figure}[htb]
\begin{center}
\epsfig{file=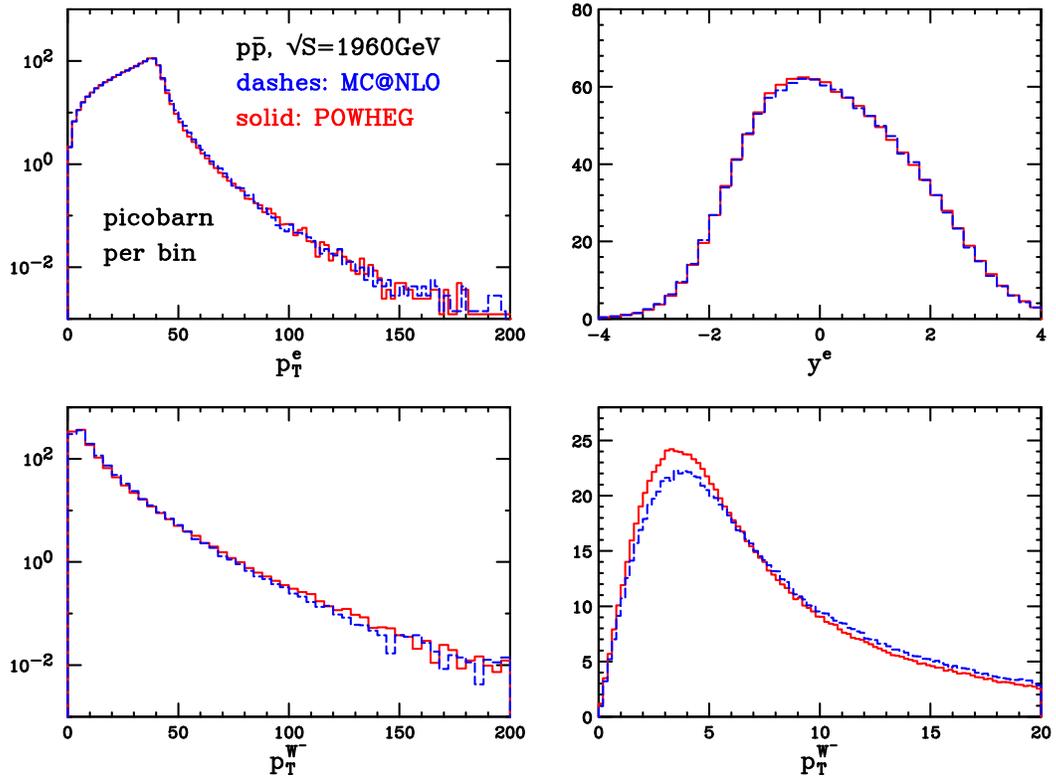,width=\figwidth}
\end{center}
\captskip
\caption{\label{fig:cmp1-wm}
Comparison of \POWHEG{} and \MCatNLO{} results for the transverse momentum
and rapidity of the lepton coming from the decay of the $W^-$ boson and for
the transverse momentum of the $W^-$, as reconstructed from its decay
product.}
\end{figure}

\begin{figure}[htb]
\begin{center}
\epsfig{file=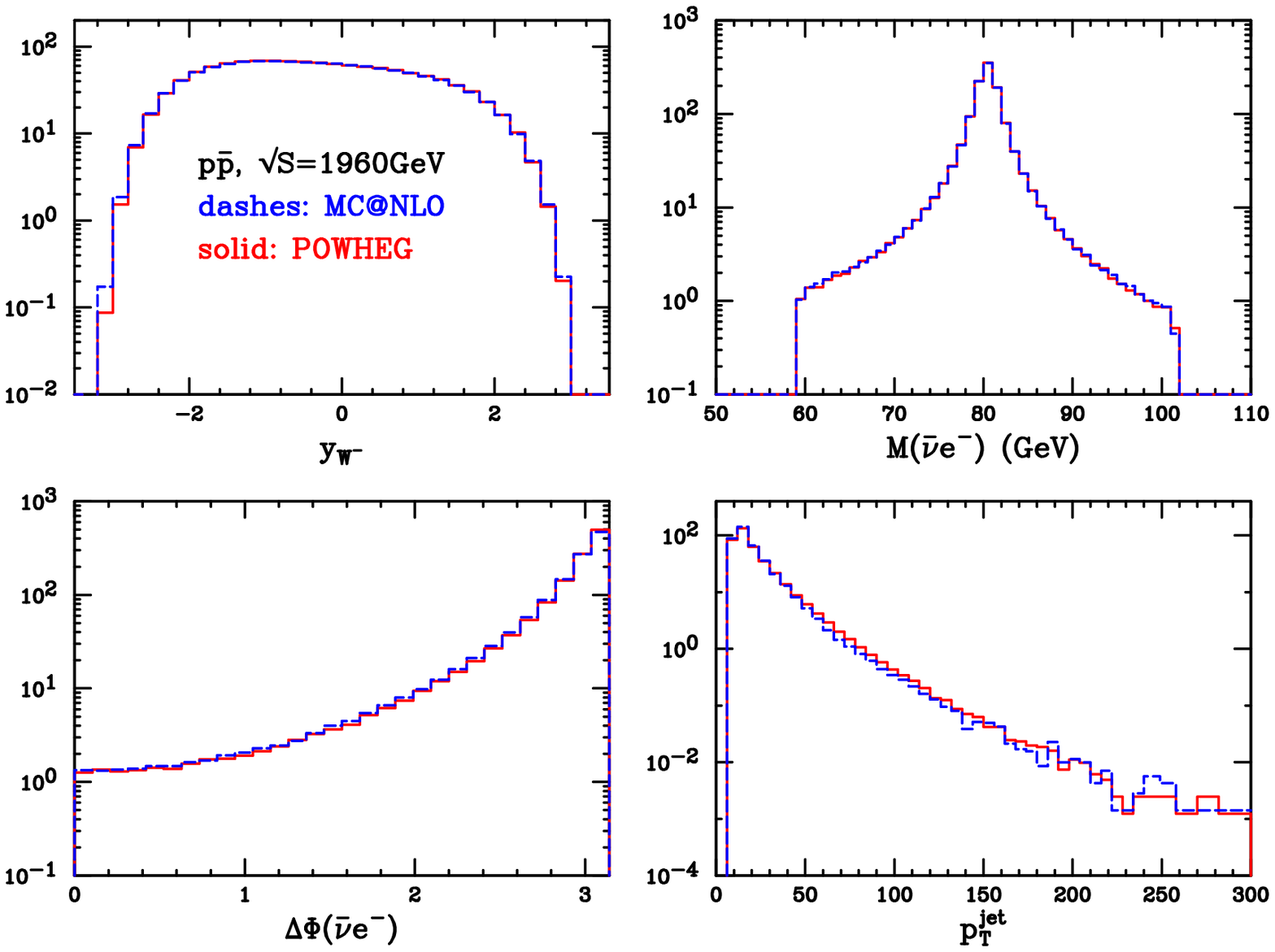,width=\figwidth}
\end{center}
\captskip
\caption{\label{fig:cmp2-wm}
Comparison of \POWHEG{} and \MCatNLO{} for the reconstructed $W^-$ rapidity,
its invariant mass, the lepton-pair azimuthal distance and the transverse
momentum of the reconstructed jet, above a 10~GeV minimum value.}
\end{figure}

\begin{figure}[htb]
\begin{center}
\epsfig{file=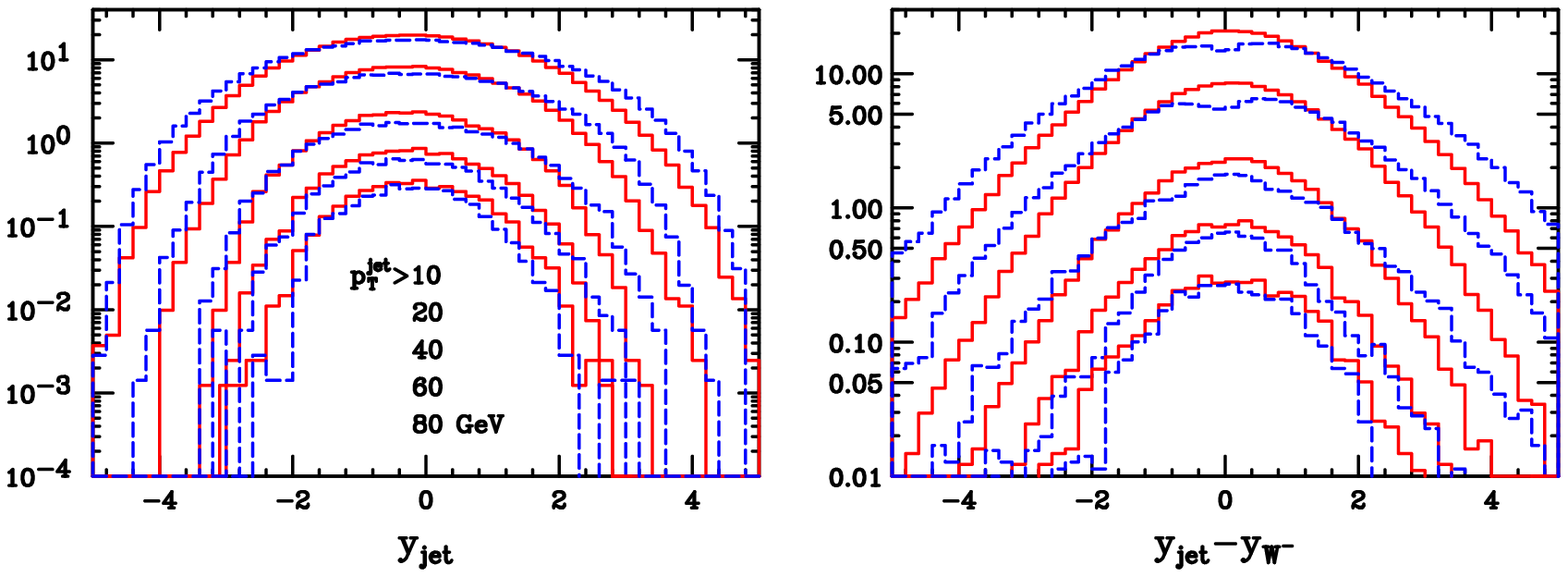,width=\figwidth}
\end{center}
\captskip
\caption{\label{fig:cmp3-wm}
Rapidity distribution of the hardest jet with different transverse-momentum
cuts, and the rapidity distance between the hardest jet and the reconstructed
$W^-$ boson for \POWHEG{} and \MCatNLO{} at the Tevatron.}
\end{figure}

\begin{figure}[htb]
\begin{center}
\epsfig{file=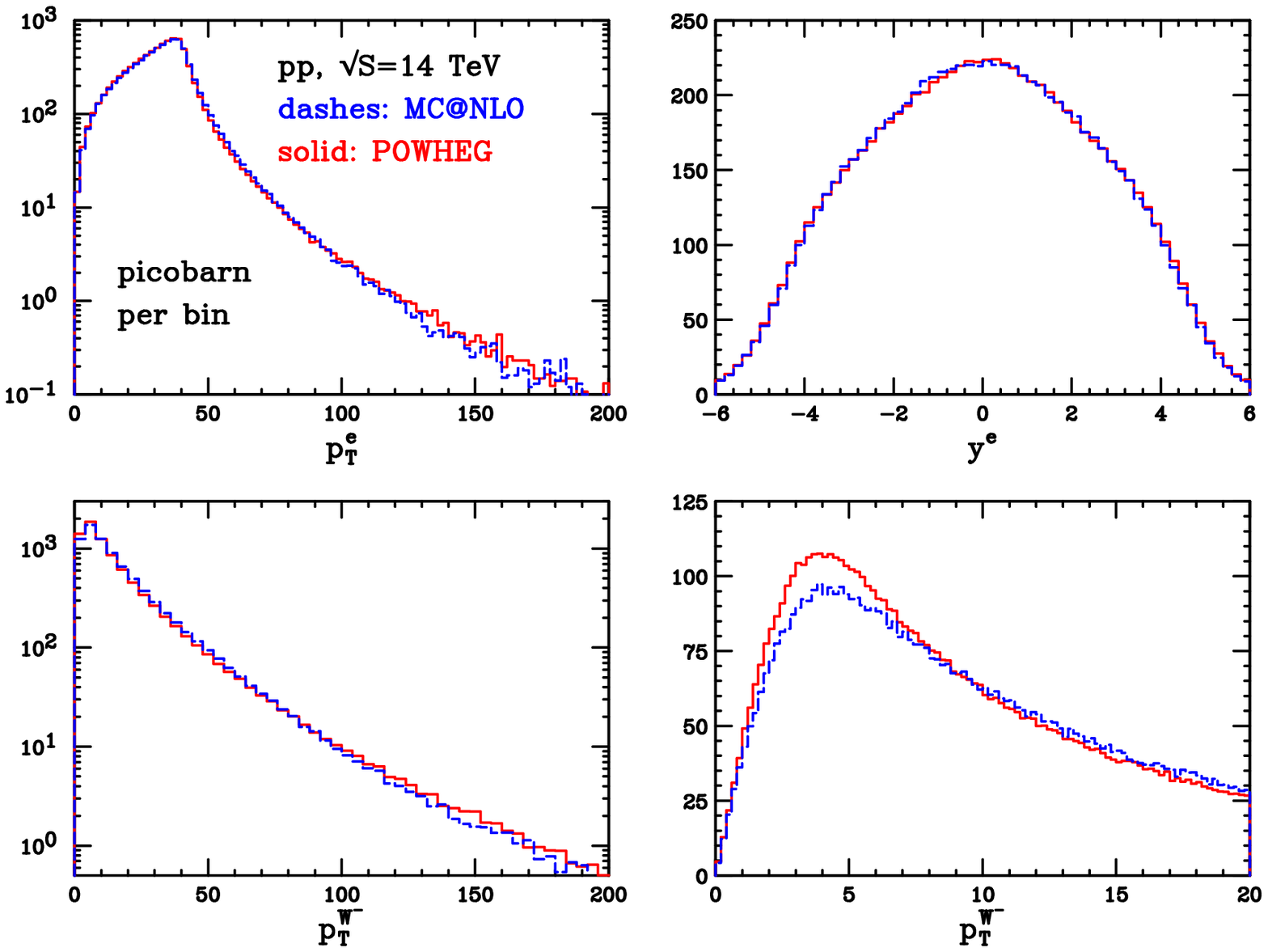,width=\figwidth}
\end{center}
\captskip
\caption{\label{fig:cmp1-wm-lhc}
Same as fig.~\ref{fig:cmp1-wm} 
for the LHC at 14~TeV.}
\end{figure}

\begin{figure}[htb]
\begin{center}
\epsfig{file=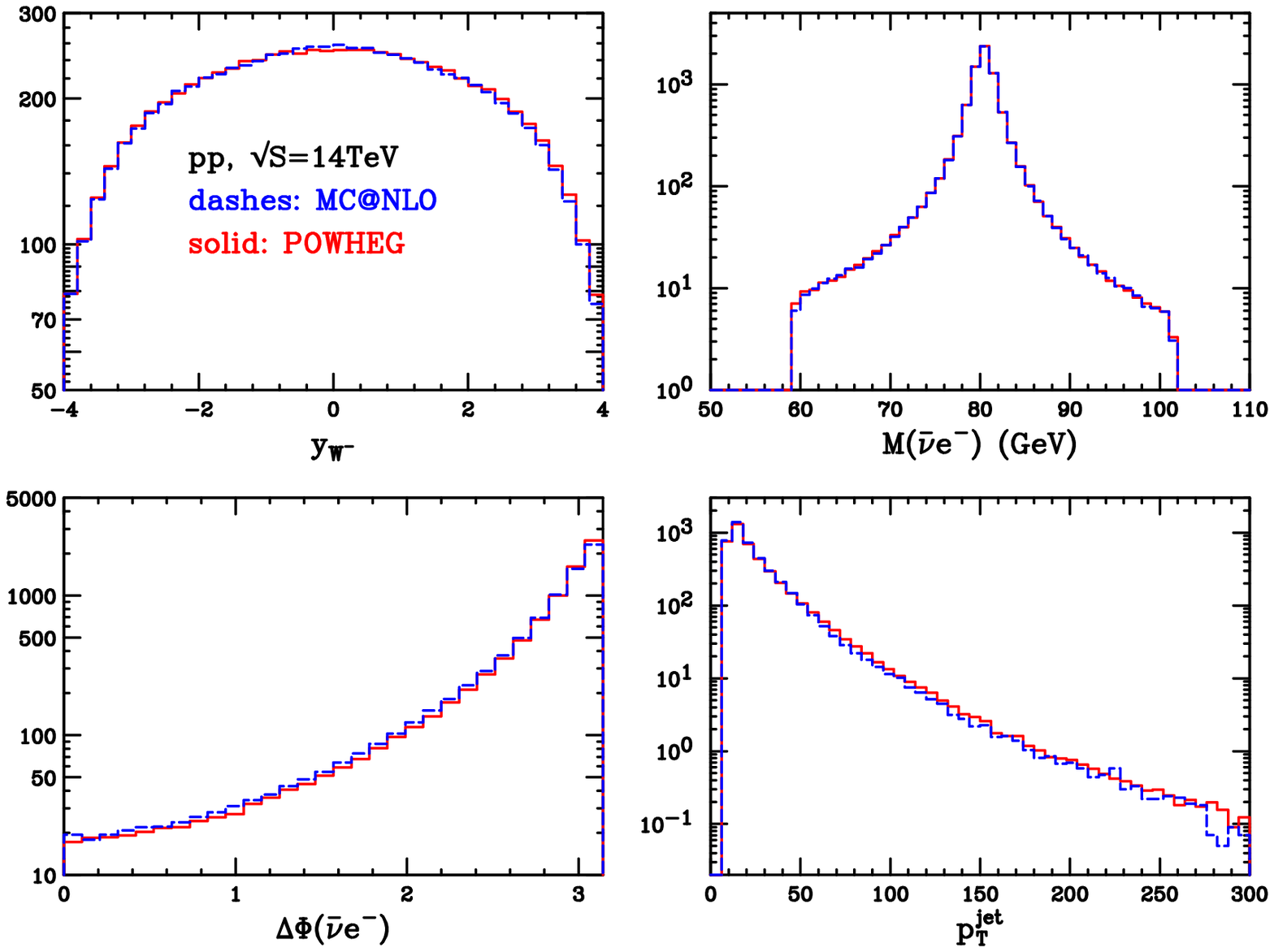,width=\figwidth}
\end{center}
\captskip
\caption{\label{fig:cmp2-wm-lhc}
Same as fig.~\ref{fig:cmp2-wm} 
for the LHC at 14~TeV.}
\end{figure}

\begin{figure}[htb]
\begin{center}
\epsfig{file=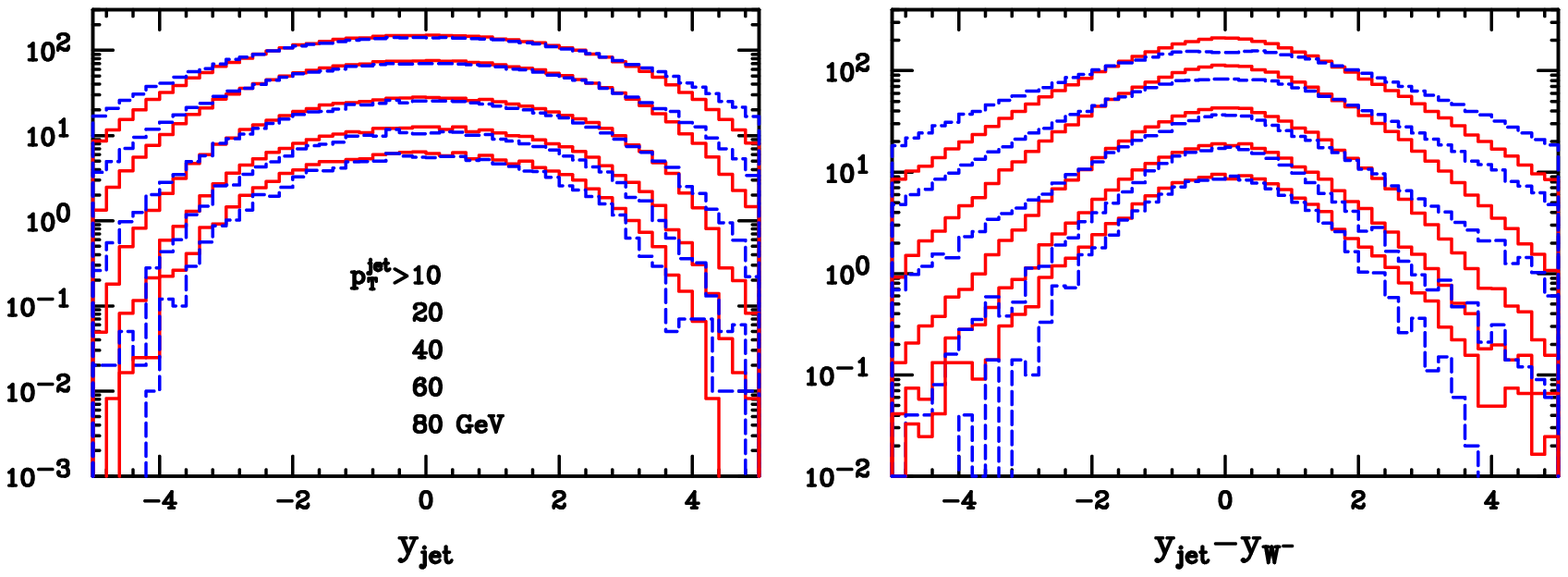,width=\figwidth}
\end{center}
\captskip
\caption{\label{fig:cmp3-wm-lhc}
Same as fig.~\ref{fig:cmp3-wm}
at the LHC at 14~TeV.}
\end{figure}

\begin{figure}[htb] %
\begin{center}
\epsfig{file=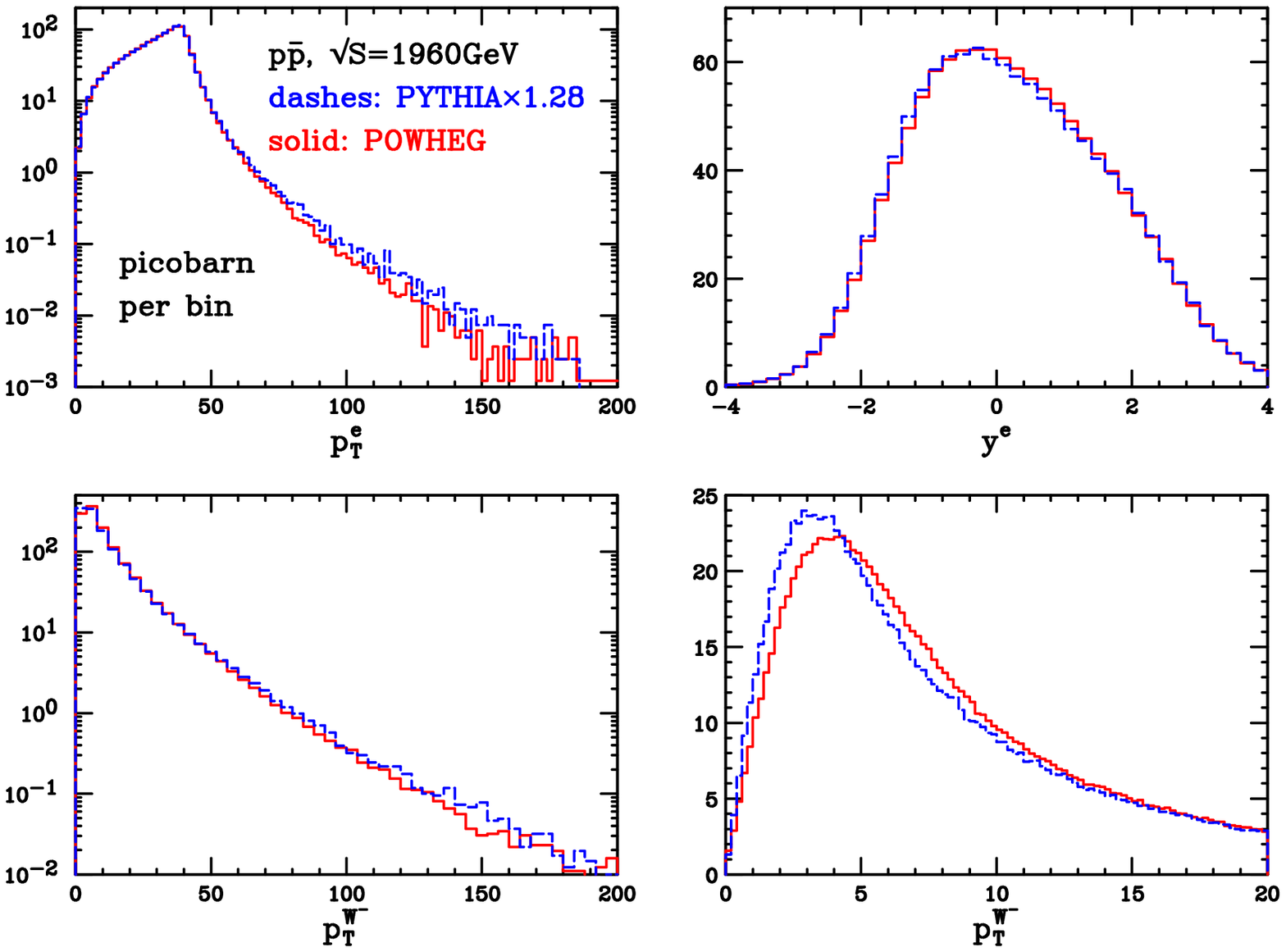,width=\figwidth}
\end{center}
\captskip
\caption{\label{fig:cmp1-wm-py}
Same as fig.~\ref{fig:cmp1-wm}
for a \PYTHIA{} and \POWHEG{} comparison at the Tevatron.}
\end{figure}

\begin{figure}[htb] %
\begin{center}
\epsfig{file=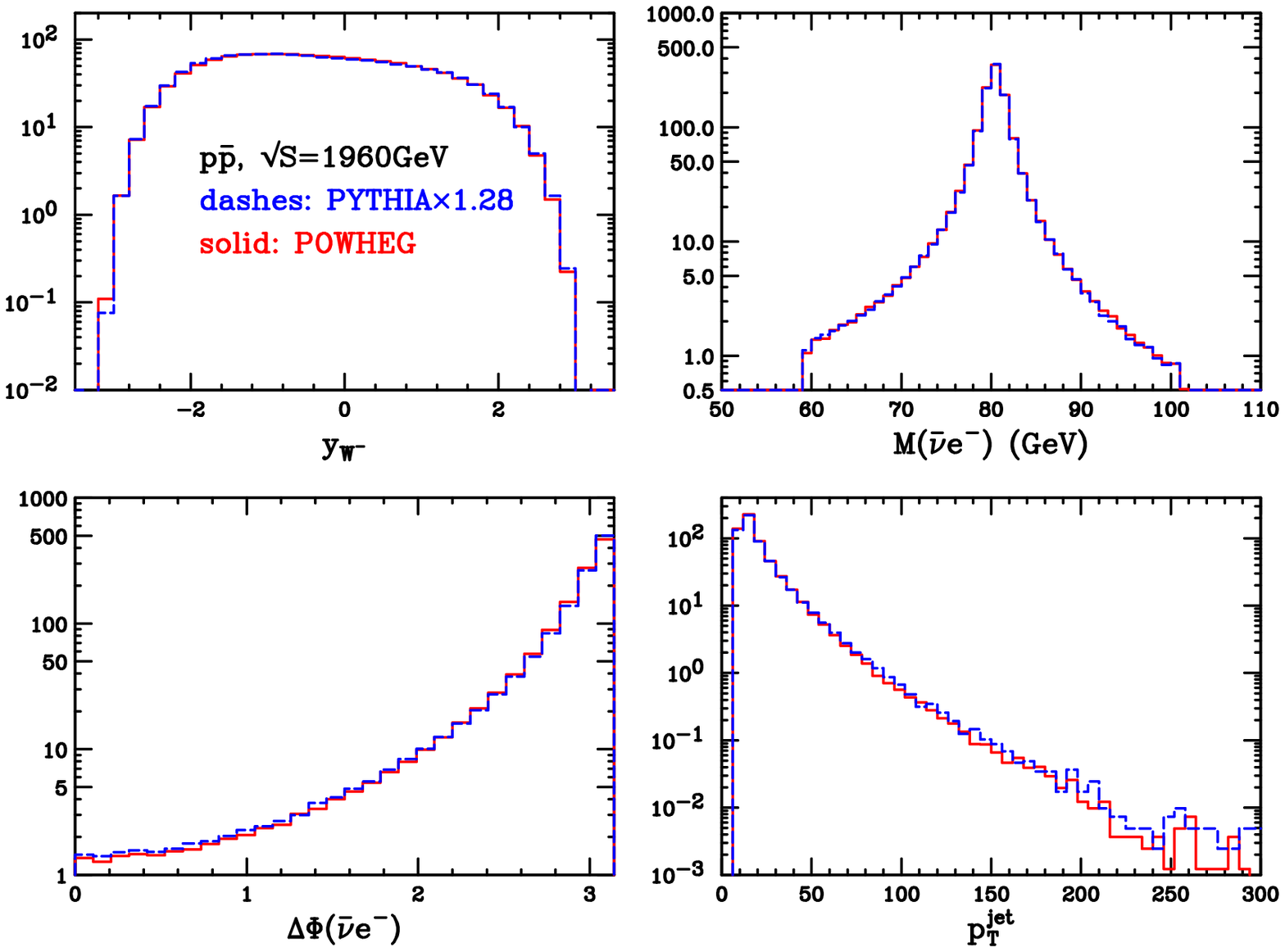,width=\figwidth}
\end{center}
\captskip
\caption{\label{fig:cmp2-wm-py}
Same as fig.~\ref{fig:cmp2-wm}
for a \PYTHIA{} and \POWHEG{} comparison at the Tevatron.}
\end{figure}

\begin{figure}[htb] %
\begin{center}
\epsfig{file=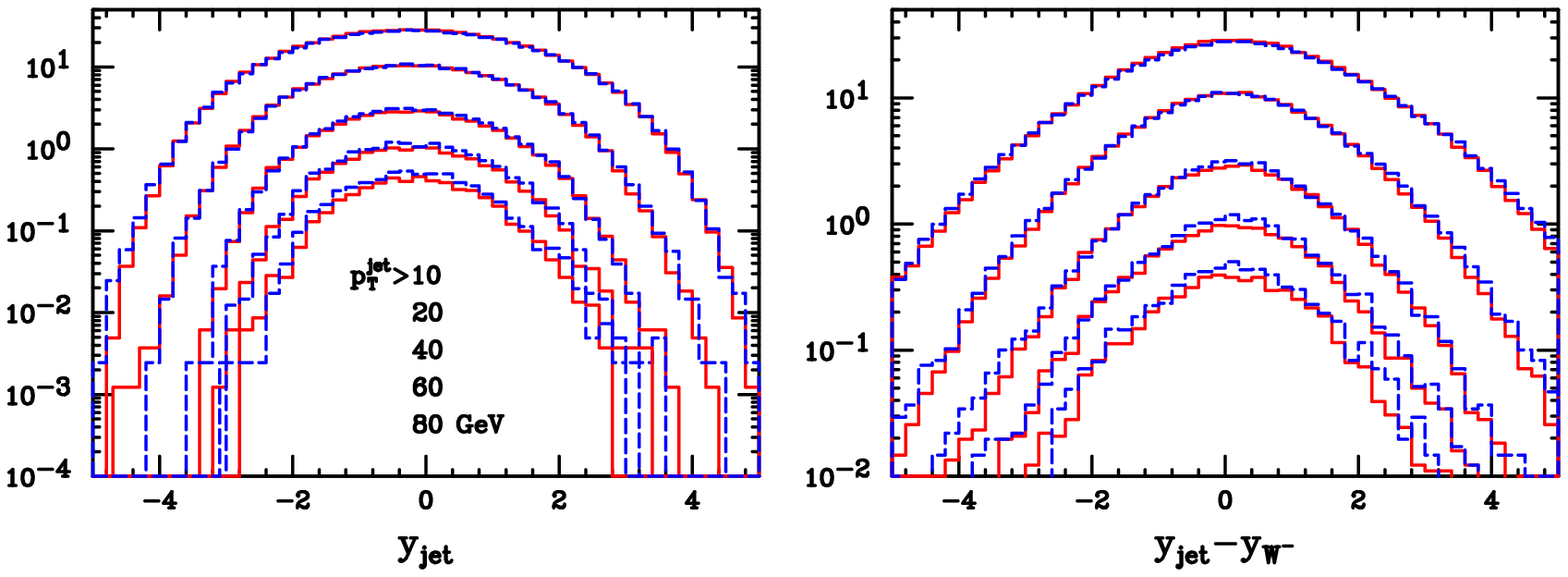,width=\figwidth}
\end{center}
\captskip
\caption{\label{fig:cmp3-wm-py} %
Same as fig.~\ref{fig:cmp3-wm}
for a \PYTHIA{} and \POWHEG{} comparison at the Tevatron.}
\end{figure}

\begin{figure}[htb] %
\begin{center}
\epsfig{file=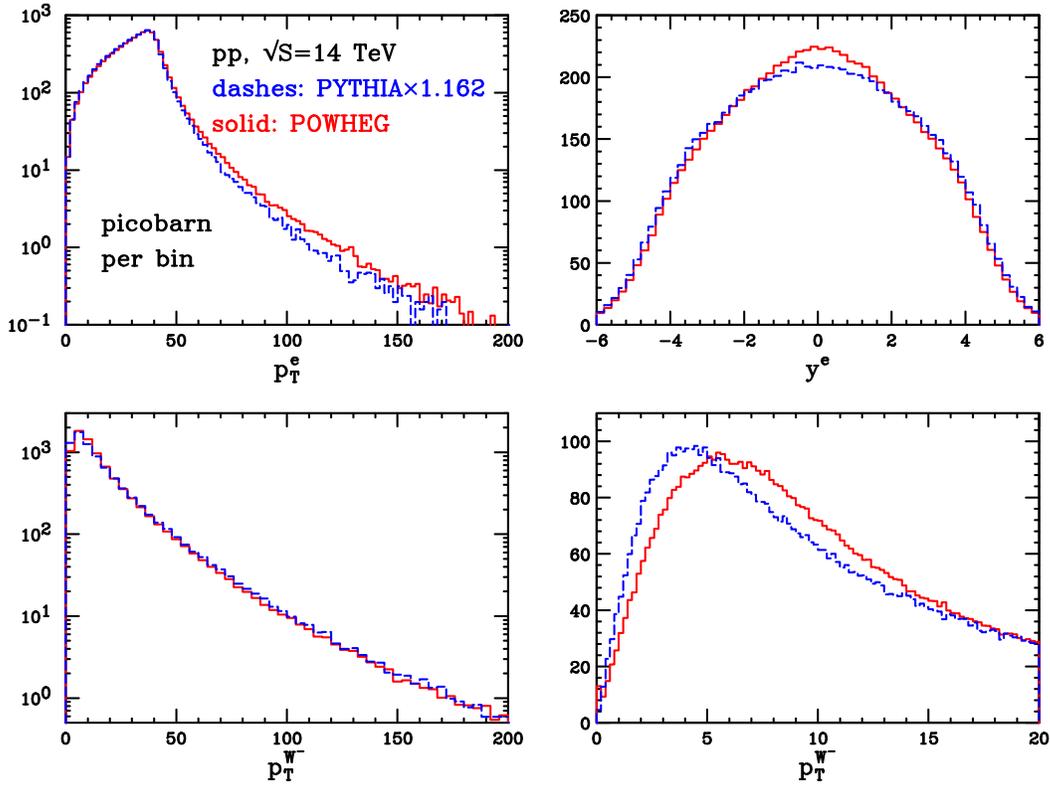,width=\figwidth}
\end{center}
\captskip
\caption{\label{fig:cmp1-wm-lhc-py} %
Same as fig.~\ref{fig:cmp1-wm}
for a \PYTHIA{} and \POWHEG{} comparison at the LHC.}
\end{figure}

\begin{figure}[htb] %
\begin{center}
\epsfig{file=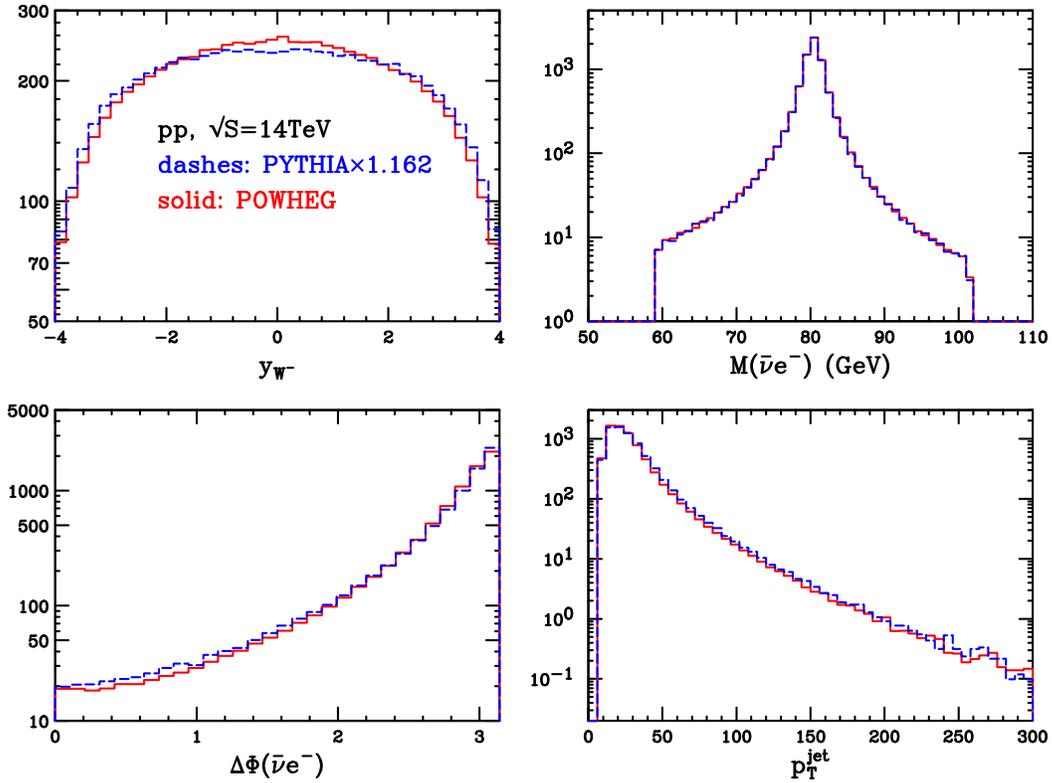,width=\figwidth}
\end{center}
\captskip
\caption{\label{fig:cmp2-wm-lhc-py} %
Same as fig.~\ref{fig:cmp2-wm}
for a \PYTHIA{} and \POWHEG{} comparison at the LHC.}
\end{figure}

\begin{figure}[htb] %
\begin{center}
\epsfig{file=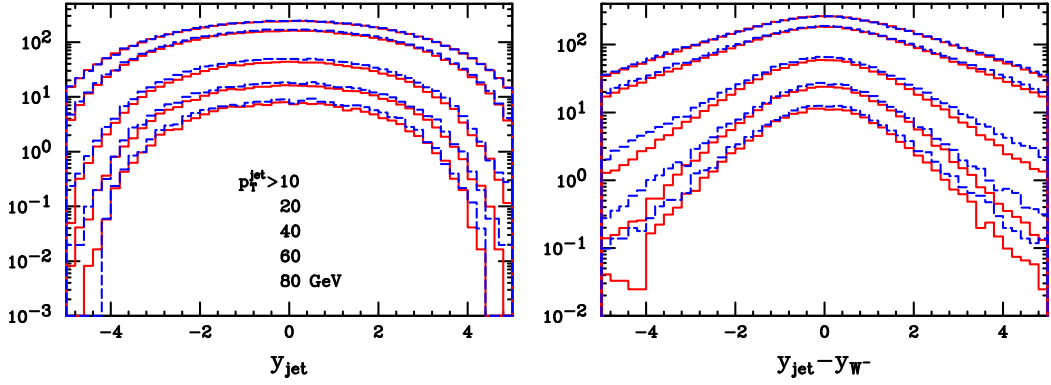,width=\figwidth}
\end{center}
\captskip
\caption{\label{fig:cmp3-wm-lhc-py}  %
Same as fig.~\ref{fig:cmp3-wm}
for a \PYTHIA{} and \POWHEG{} comparison at the LHC.}
\end{figure}

\begin{figure}[htb] %
\begin{center}
\epsfig{file=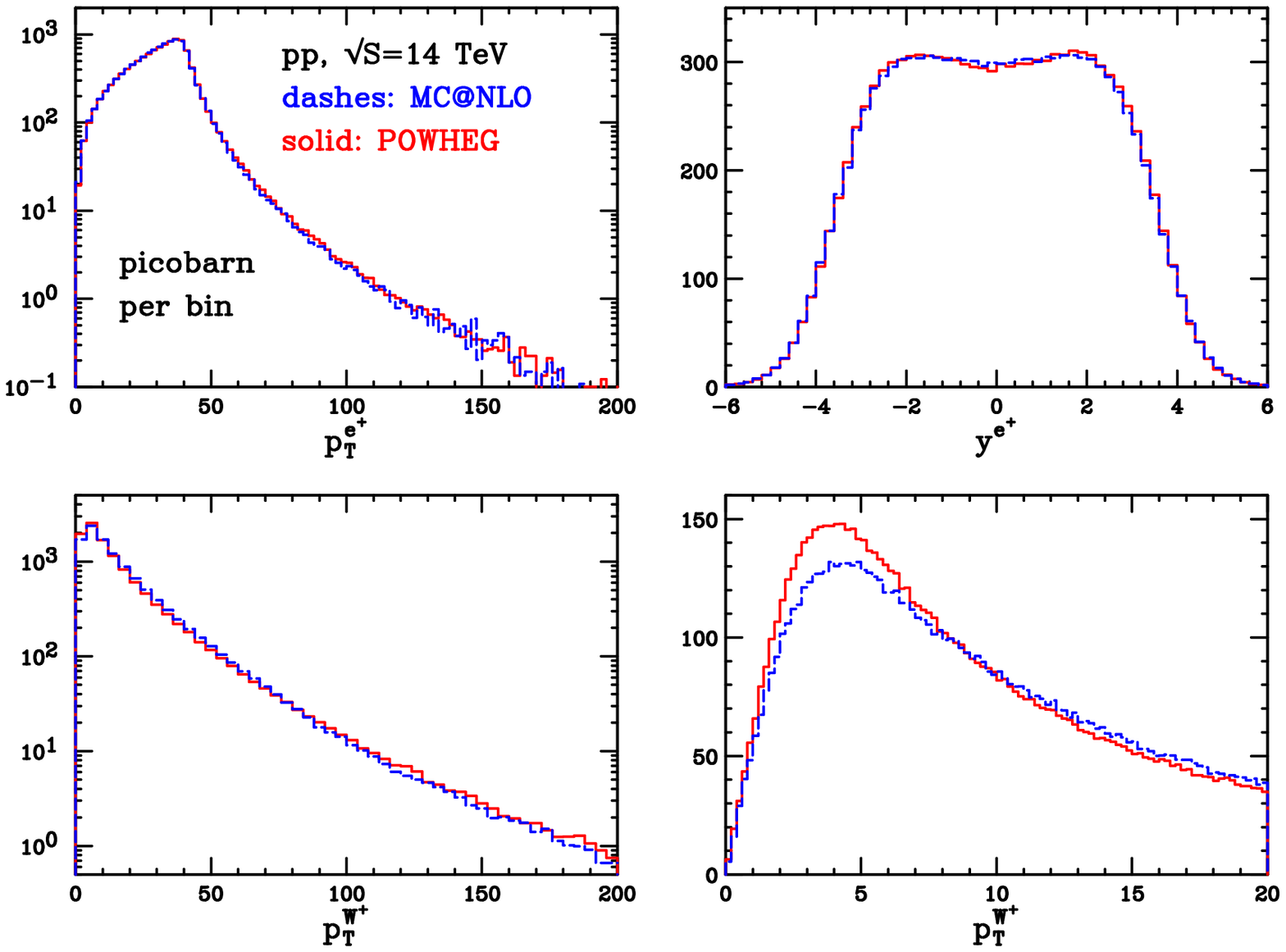,width=\figwidth}
\end{center}
\captskip
\caption{\label{fig:cmp1-wp-lhc}  %
Same as fig.~\ref{fig:cmp1-wm-lhc} for $W^+$ production  at the LHC.}
\end{figure}

\begin{figure}[htb] %
\begin{center}
\epsfig{file=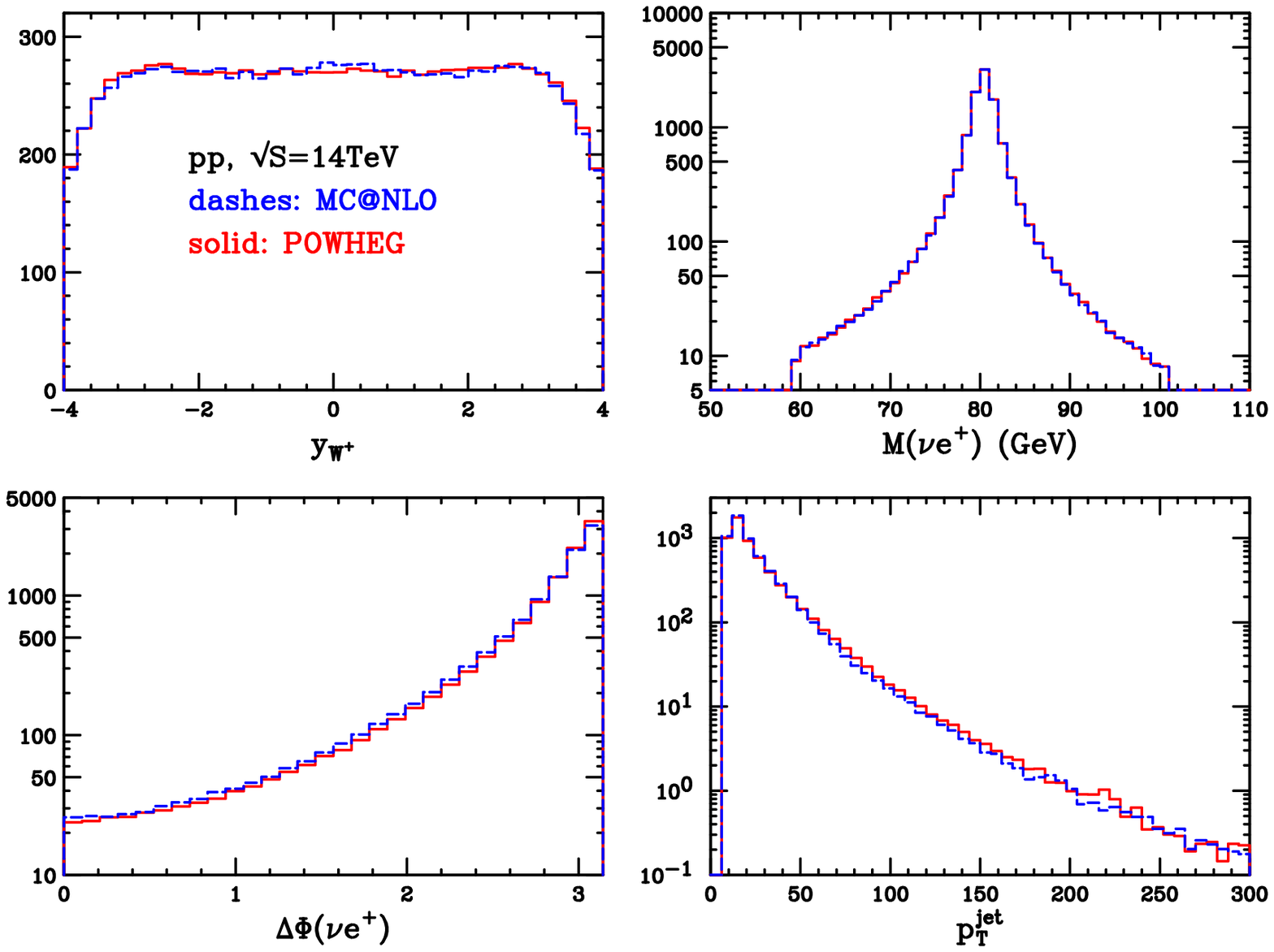,width=\figwidth}
\end{center}
\captskip
\caption{\label{fig:cmp2-wp-lhc}  %
Same as fig.~\ref{fig:cmp2-wm-lhc} for $W^+$ production  at the LHC.}
\end{figure}

\begin{figure}[htb] %
\begin{center}
\epsfig{file=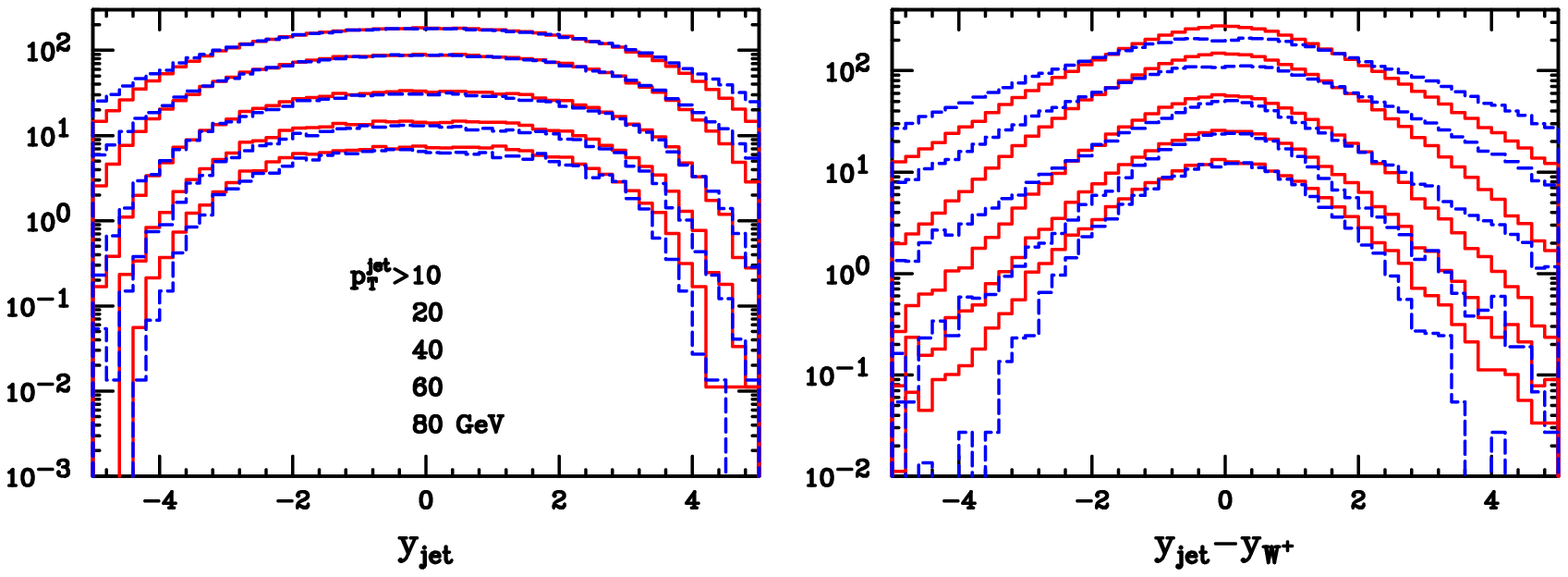,width=\figwidth}
\end{center}
\captskip
\caption{\label{fig:cmp3-wp-lhc}  %
Same as fig.~\ref{fig:cmp3-wm-lhc} for $W^+$ production  at the LHC.}
\end{figure}

\begin{figure}[htb] %
\begin{center}
\epsfig{file=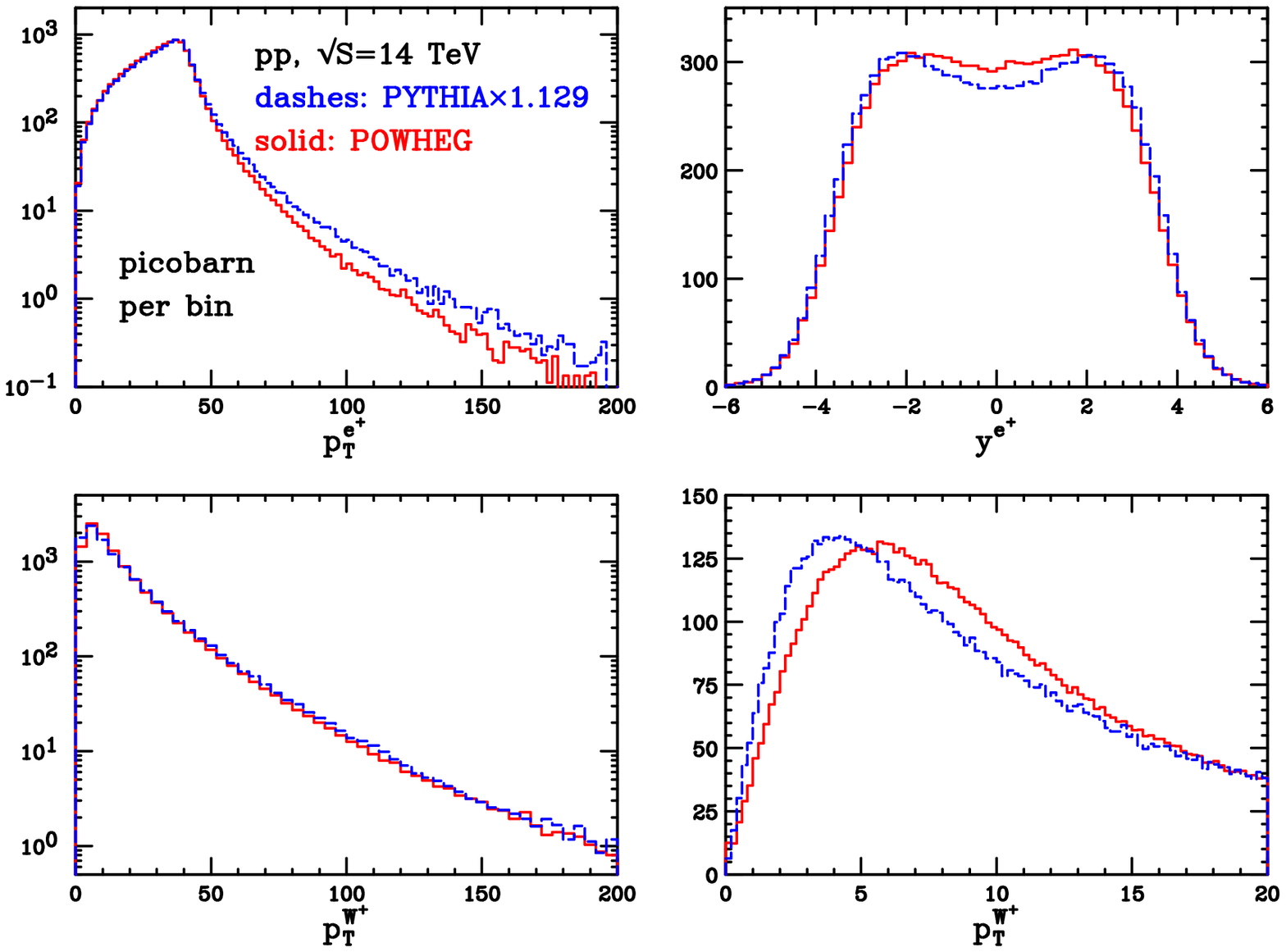,width=\figwidth}
\end{center}
\captskip
\caption{\label{fig:cmp1-wp-lhc-py}  %
Same as fig.~\ref{fig:cmp1-wm-lhc-py} for $W^+$ production  at the LHC.}
\end{figure}

\begin{figure}[htb] %
\begin{center}
\epsfig{file=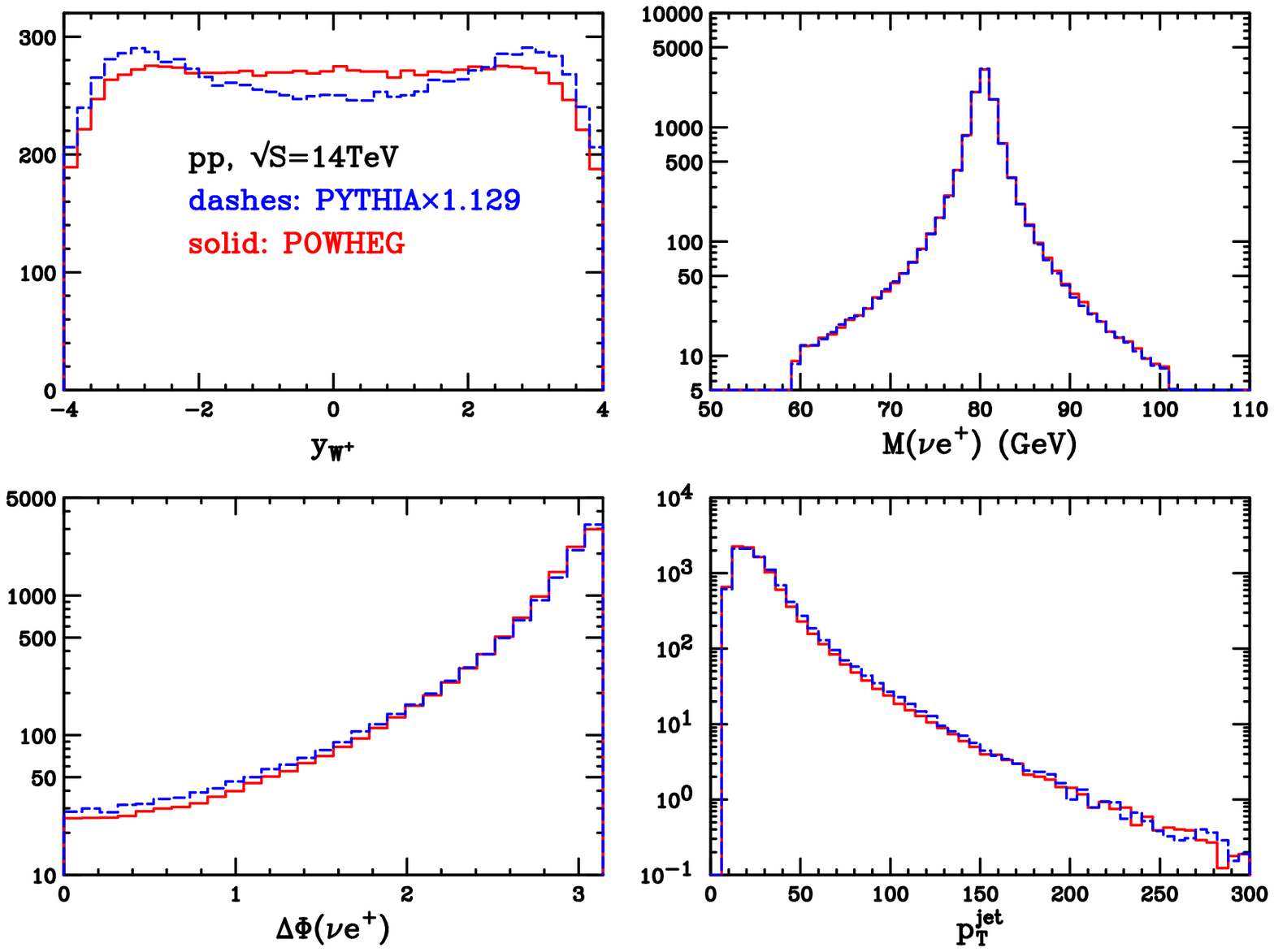,width=\figwidth}
\end{center}
\captskip
\caption{\label{fig:cmp2-wp-lhc-py}  %
Same as fig.~\ref{fig:cmp2-wm-lhc-py} for $W^+$ production  at the LHC.}
\end{figure}

\begin{figure}[htb] %
\begin{center}
\epsfig{file=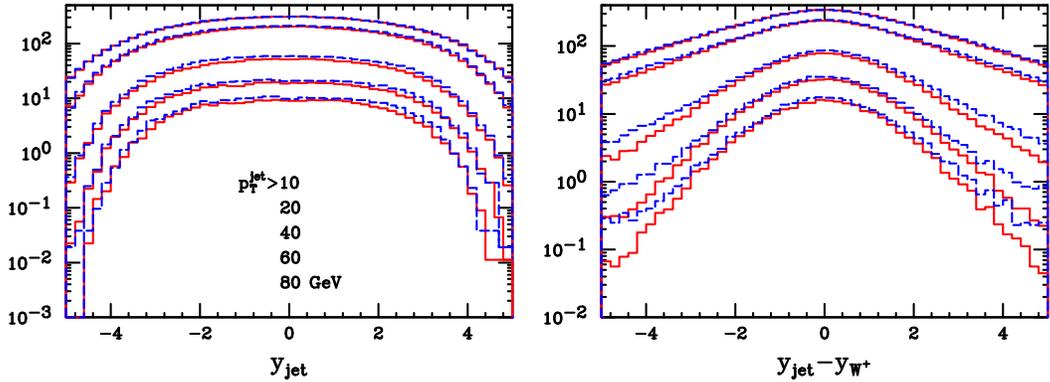,width=\figwidth}
\end{center}
\captskip
\caption{\label{fig:cmp3-wp-lhc-py}  %
Same as fig.~\ref{fig:cmp3-wm-lhc-py} for $W^+$ production  at the LHC.}
\end{figure}

All results presented so far are relative to $Z$ boson production. In the
case of $W$ production we find similar features and the comparison between
\MCatNLO{} and \PYTHIA{} presents very similar characteristics.  For the sake
of completeness, we present in fig.~\ref{fig:cmp1-wm}
through~\ref{fig:cmp3-wp-lhc-py} plots of observables for $W^-$ production at
the Tevatron, and $W^-$ and $W^+$ production at the LHC, comparing again the
\POWHEG{} output with \MCatNLO{} and \PYTHIA{}, and the observables for $W^+$
production at the LHC. We find again that \MCatNLO{} displays dips in the
rapidity distribution of the hardest jet at Tevatron energy.  The comparison
of the transverse-momentum distribution of the $W$ shows the same differences
found in the $Z$ case. Furthermore, the rapidity distribution of the $W^\pm$
at the LHC differs in \PYTHIA{}, showing a very marked difference in the
$W^+$ case (see fig.~\ref{fig:cmp2-wp-lhc-py}), probably (as in the $Z$ case)
a consequence of the different pdf set.
\clearpage

\section{Conclusions}
\label{sec:conc}
In this paper we have reported on a complete implementation of vector-boson
production at NLO in the \POWHEG{} framework. The calculation was performed
within the Catani-Seymour~\cite{Catani:1997vz} dipole approach, and thus this
is the first \POWHEG{} implementation within the Catani-Seymour framework at
a hadronic collider.  We have found that, at variance with what was proposed
in sec.~7.3 of ref.~\cite{Frixione:2007vw}, it is better to define the
transverse momentum as the true transverse momentum for the initial-state
singular region. Furthermore, we have shown how to perform a \POWHEG{}
implementation when the Born term vanishes.

The results of our work have been compare extensively with \MCatNLO{} and
\PYTHIA{}. The \PYTHIA{} result, rescaled to the full NLO cross section, is
in good agreement with \POWHEG{}, except for differences in the rapidity
distribution of the vector boson, that may be ascribed to the use of a LO
parton density in \PYTHIA. The \MCatNLO{} result is in fair agreement with
\POWHEG, except for the distribution of the hardest jet in the process, the
\MCatNLO{} distribution being generally wider. Furthermore, we have also
examined the distributions in the difference of the hardest jet and the
vector-boson rapidity. We have found that these distributions exhibit dips at
zero rapidity. We have also examined analogous distributions for $ZZ$ and
$t\bar{t}$ production, and again found dips for these distributions, that seem
to be a general feature of the \MCatNLO{} approach. We also remark that no
other approaches show dips of this kind~\cite{Alwall:2007fs}.

The computer code for the \POWHEG{} implementations presented here
is available, together with the manual, at the site\newline
\centerline{{\tt http://moby.mib.infn.it/\~{}nason/POWHEG\ }.}

\section{Acknowledgments}
We wish to thank S.~Frixione, K.~Hamilton, T.~Sj\"ostrand and B.~Webber for
useful discussions.  This research was supported in part by the National
Science Foundation under Grant No. PHY05-51164.

\appendix
\section{Upper bounding function}%
\label{app:uboundfun}
We call $\Delta_U(p_T^2)$ the Sudakov form factor obtained with the upper
bounding function of eq.~(\ref{eq:uboundf}). Using the definitions 
of eqs.~(\ref{eq:dphirad}) and~(\ref{eq:ktsq})
\beqn
 d \Phi_{\tmop{rad}} &=& \frac{M^2}{16 \pi^2} \, \frac{d \phi}{2 \pi}
  \,d v \, \frac{d x}{x^2} \,\theta (v)\, \theta \!\left( 1 - \frac{v}{1 - x}
  \right) \theta (x (1 - x)) \, \theta (x - \bar{x}_{\splus})\\
 k_T^2 &=& \frac{M^2}{x} (1 - x - v) \,v\;,
\eeqn
we write
\begin{eqnarray}
  \frac{\log \Delta_U (p_T^2)}{- N} & = & \int_{\bar{x}}^1 \frac{d x}{x^2}
  \int_0^{1 - x} d v \, \frac{\alpha_s (k_T^2)}{2 v}  \frac{x^2}{1 - x - v}
 \, \theta\!\(k_T^2 - p_T^2\)\nonumber \\
  & = & \int_{\bar{x}}^1 \frac{d x}{x} \int_0^{1 - x} d v \,\frac{\alpha_s
  (k_T^2)}{2} \, \frac{M^2}{k_T^2}\, \theta\!\(k_T^2 - p_T^2\)
 \nonumber\\\nonumber 
  & = & \int_{p_T^2}^{\infty} \frac{d k_T^2}{k_T^2}\,  \frac{\alpha_s
  (k_T^2)}{2} \int_0^1 d v \int_{\bar{x}}^1 \frac{d x}{x}\, \theta (1 - x - v)
  \,M^2\, \delta \!\left( \frac{M^2}{x} (1 - x - v) v - k_T^2 \right)\,,
\end{eqnarray}
where, for ease of notation, we have dropped the $\splusminus$ and $q\bar{q}$
labels on $N$ and $\bar{x}$.
We perform the $x$ integration using the $\delta$ function
\begin{equation}
  \int \frac{d x}{x} \,M^2\, \delta\!\left( \frac{M^2}{x} (1 - x - v) v - k_T^2
  \right) = \frac{1}{k_T^2 / M^2 + v},\qquad\quad
 x = \frac{M^2 v (1 - v)}{k_T^2 + M^2 v}\;.
\end{equation}
Notice that $x < 1$, and
\begin{equation}
  \theta \!\left( 1 - v - \frac{M^2 v (1 - v)}{k_T^2 + M^2 v} \right) = \theta
 \! \left( 1 - v\frac{k_T^2+ M^2 }{k_T^2 + M^2 v} \right) = 1 .
\end{equation}
The only remaining condition on $x$ is $x \geqslant \bar{x}$. We thus get
\begin{equation}
  \frac{\log \Delta_U (p_T^2)}{- N} = \int_{p_T^2}^{\infty}\, \frac{d
  k_T^2}{k_T^2}  \frac{\alpha_s (k_T^2)}{2} \int_0^1 \frac{d v}{k_T^2 / M^2 +
  v} \,\theta\! \left( \frac{M^2 v (1 - v)}{k_T^2 + M^2 v} - \bar{x} \right) .
\end{equation}
We must find the conditions implied by the theta function upon $v$. For
\begin{equation}
  k_T^2 < k_{T \max}^2 = \frac{M^2 (1 - \bar{x})^2}{4 \bar{x}}\,,
\end{equation}
the $\theta$ function is satisfied if $v_- < v < v_+$, where
\begin{equation}
  v_{\pm} = \frac{1 - \bar{x} \pm \sqrt{(1 - \bar{x})^2 - 4 \,\bar{x}\,
  \frac{k_T^2}{M^2}}}{2} \,.
\end{equation}
We thus have
\begin{equation}
  \frac{\log \Delta_U (p_T^2)}{- N} = \int_{p_T^2}^{k_{T \max}^2} \frac{d
  k_T^2}{k_T^2}\,  \frac{\alpha_s (k_T^2)}{2}\, \log \frac{\frac{k_T^2}{M^2} +
  v_+}{\frac{k_T^2}{M^2} + v_-}\,.
\end{equation}
The $k_T^2$ integral is still too complex to be performed analytically. We
thus resort another time to the veto method, by finding an upper bound to the
integrand. We have
\begin{equation}
  \frac{\frac{k_T^2}{M^2} + v_+}{\frac{k_T^2}{M^2} + v_-} \leqslant
  \frac{\frac{k_T^2}{M^2} + 1}{\frac{k_T^2}{M^2}} = \frac{M^2}{k_{T^{}}^2} + 1
  \leqslant \frac{M^2}{k_{T^{}}^2} + \frac{k_{T \max}^2}{k_{T^{}}^2} =
  \frac{M^2 (1 + \bar{x})^2}{4 \,\bar{x} \,k_T^2} .
\end{equation}
We thus define
\begin{equation}
  q^2 = \frac{M^2 (1 + \bar{x})^2}{4 \, \bar{x}\, k_T^2},
\end{equation}
and introduce a new Sudakov form factor
\begin{equation}
\label{eq:doubleupper}
  \frac{\log \tilde{\Delta}_U (p_T^2)}{- N} = \int_{p_T^2}^{k_{T \max}^2}
  \frac{d k_T^2}{k_T^2} \, \frac{\alpha_U (k_T^2)}{2} \,\log
  \frac{q^2}{k_T^2}, 
\end{equation}
where $\alpha_U(k_T^2)$ has the form of the one-loop running coupling
constant
\begin{equation}
  \alpha_U (k_T^2) = \frac{1}{b\log \frac{k_T^2}{\Lambda_U^2}},
\end{equation}
and is required to satisfy the bound $\alpha_U (k_T^2) \geqslant \alpha_s
(k_T^2)$ in the allowed range for $k_T^2$. The integral in
eq.~(\ref{eq:doubleupper}) is now easily performed, and we get 
\begin{equation}
  \tilde{\Delta}_U (p_T^2) = \exp \left\{ - \frac{N}{2 b} \left[ \log
  \frac{q^2}{\Lambda_U^2} \log \frac{\log \frac{k_{T
  \max}^2}{\Lambda_U^2}}{\log \frac{p_T^2}{\Lambda_U^2}} - \log \frac{k_{T
  \max}^2}{p_T^2} \right] \right\} .
\end{equation}
The generation of the radiation variables is then performed starting with
$\tilde{\Delta}_U (p_T^2)$, using the
veto procedure to obtain the $\text{$\Delta_U (p_T^2)$}$ distribution.
Further vetoing is then used to obtain the correct $R/B$ generated
distribution.

\bibliography{paper}
\end{document}